%% file: paper.tex
\documentclass[journal]{IEEEtran}
\usepackage{array,graphicx,cite,epsfig,amssymb,amsfonts}
\usepackage{bbm,amsmath,epstopdf,algorithm,algorithmic,subfigure, multirow}
\usepackage{comment}
\usepackage{color}
\usepackage{cases}
\usepackage{slashbox}

\newcommand{\be}{\begin{equation}}
\newcommand{\ee}{\end{equation}}
\def\bee#1\eee{\begin{align}#1\end{align}}
\newcommand{\bse}{\begin{subequations}}
\newcommand{\ese}{\end{subequations}}
\newcommand{\nnb}{\nonumber}

\newtheorem{theorem}{\textbf{Theorem}}

\ifodd 0
\else

\fi

\begin{document}

\title{Reinforcement Learning Random Access for Delay-Constrained Heterogeneous Wireless Networks: A Two-User Case}

\author{Danzhou~Wu,
		Lei~Deng,~\IEEEmembership{Member,~IEEE,}
        Zilong~Liu,~\IEEEmembership{Senior Member,~IEEE,}
        Yijin~Zhang,~\IEEEmembership{Senior Member,~IEEE,}
        and~Yunghsiang~S.~Han,~\IEEEmembership{Fellow,~IEEE}
\thanks{
This work was supported in part by the National Natural Science Foundation of China under Grants 61902256, 62071236, and 61671007,
in part by Tencent ``Rhinoceros Birds"-Scientific Research Foundation for Young Teachers of Shenzhen University,
and in part by the Fundamental Research Funds for the Central Universities of China under Grant 30920021127.
}
\thanks{D.~Wu and L.~Deng are with College of Electronics and Information Engineering,
Shenzhen University, Shenzhen 518060, China
also with the Shenzhen Key Laboratory of Digital Creative Technology, Shenzhen 518060, China, and also with the Guangdong Province Engineering Laboratory for Digital Creative Technology, Shenzhen 518060, China
(e-mail: wudanzhou2019@email.szu.edu.cn, ldeng@szu.edu.cn).}
\thanks{Z.~Liu is with School of Computer Science and Electronic Engineering, University of Essex, Colchester CO4 3SQ, U.K. (E-mail:
zilong.liu@essex.ac.uk).}
\thanks{Y.~Zhang is with School of Electronic and Optical Engineering, Nanjing University of Science and Technology, Nanjing 210094, China (e-mail:
yijin.zhang@gmail.com).}
\thanks{Y.~S.~Han is with Shenzhen Institute for Advanced Study, University of Electronic Science and Technology of China, Shenzhen 518110, China  (e-mail: yunghsiangh@gmail.com).}
}


\maketitle

\input{abstract.tex}

\IEEEpeerreviewmaketitle

\input{introduction.tex}

\input{model.tex}
\input{Upperbound.tex}

\input{TSRA.tex}

\input{simulation.tex}

\input{conclusion.tex}

\bibliographystyle{IEEEtran}
\bibliography{ref}

\input{appendix.tex}

\vfill



%

\end{document}

%% file: abstract.tex
\begin{abstract}
In this paper, we investigate the random access problem for a delay-constrained  heterogeneous wireless network.
As a first attempt to study this new problem, we consider a network with two users who deliver delay-constrained
traffic to an access point (AP) via a common unreliable collision wireless channel.
By assuming that one user (called user 1) adopts ALOHA, we aim to optimize 
the random access scheme of the other user (called user 2).
The most intriguing part of this problem is that user 2 does not know the information
of user 1 but needs to  maximize the system timely throughput.
Such a paradigm of collaboratively sharing spectrum  is envisioned by DARPA to better dynamically match the supply and demand
in future networks \cite{SC2,tilghman2019will}. We first propose a Markov Decision Process (MDP) formulation
to derive a model-based upper bound so as to quantify the performance gap of any designed schemes.
We then utilize reinforcement learning (RL) to design an R-learning-based \cite{schwartz1993reinforcement, singh1994reinforcement,sutton2018reinforcement} random access scheme, called TSRA.
We carry out extensive simulations to show that TSRA achieves close-to-upper-bound performance and better performance than
the existing baseline DLMA \cite{yiding2019deep}, which is our counterpart scheme for delay-unconstrained heterogeneous wireless network.
All source code is publicly available in \texttt{https://github.com/DanzhouWu/TSRA}.
\end{abstract}


%% file: introduction.tex
\section{Introduction}\label{sec:introduction}
\IEEEPARstart{C}{ommunication} is shifting its role from connecting people to networking everything in various vertical domains.
Toward that end, hard delay constraint is one of the most important communication requirements in many vertical applications,
such as factory automation, robot collaboration and control, smart grid load control, autonomous vehicles, online gaming,
virtual reality, tactile Internet, etc. \cite{ts22104,chen2020wireless, gp2015the, kkim2012cyber,lei2017timely}.
In such applications, each packet has a hard deadline:
it will expire and then be removed from the system if it has not been delivered successfully before its deadline.
For example, in virtual reality, the motion-to-photon latency is generally at most 15 ms; exceeding
this deadline will cause motion sickness and dizziness to the user \cite{elbamby2018toward}.

To support various applications in different scenarios, heterogeneous wireless networks are ubiquitous nowadays.
It is common that different networks, such as cellular, WiFi, Bluetooth, Zigbee, LoRa, NFC, etc., co-exist in an area to deliver data traffic.
Currently, spectrum is generally rigidly divided into exclusively occupied bands among different networks
to mitigate interference. This exclusively-assigning scheme, however, is hard to
satisfy the explosively increasing wireless traffic, since it is unable to dynamically match the supply and demand.
To address this issue, the Defense Advanced Research Projects Agency (DARPA)
envisions that spectrum should be dynamically and collaboratively shared by
heterogeneous wireless networks. To validate this new spectrum sharing scheme,
DARPA hosted a three-year competition, called Spectrum Collaboration Challenge (SC2),
where teams need to design clean-slate radio techniques to share spectrum with their competitors but without knowing
protocol details of competitors, \emph{with the ultimate goal of increasing overall data throughput} \cite{SC2,tilghman2019will}.
The competition has demonstrated that indeed the new collaboratively-sharing scheme can transmit far more data than the inflexible exclusively-assigning scheme.
To realize DARPA's vision, we need to re-design PHY, MAC and network layers of wireless networks.
In this paper, we only focus on the MAC layer design, in particular, on the uplink random access scheme design.

New random access schemes have been designed in heterogeneous wireless networks for delay-unconstrained communications.
Yu \emph{et al}. in \cite{yiding2019deep} introduced deep reinforcement learning (DRL) into the random access scheme design for heterogeneous wireless networking.
Their proposed scheme, called deep-reinforcement learning multiple access (DLMA), adopted feedforward neural networks (FNN) as the deep neural network.
In \cite{yiding2020non}, the authors further applied DRL into CSMA and designed a new CSMA variant, called CS-DLMA, for heterogeneous wireless networking.
As compared with DLMA, CS-DLMA adopts recurrent neural networks (RNN) for a non-uniform time-step deep Q-network (DQN) by leveraging the fact that
the time duration required for carrier sensing is smaller than the duration of data transmission. Both \cite{yiding2019deep} and \cite{yiding2020non}
assume a saturated delay-unconstrained traffic pattern.
On the other hand, some works studied random access schemes for delay-constrained communication in homogeneous wireless networks.
Deng \emph{et al.} in\cite{lei2018on} analyzed the asymptotic performance of ALOHA system for frame-synchronized delay-constrained traffic pattern.
Zhang \emph{et al.} studied the system throughput and optimal retransmission probability of ALOHA for the saturated delay-constrained traffic \cite{zhang2019achieving} .
\cite{campolo2011modeling} analyzed $p$-persistent CSMA for broadcasting delay-constrained traffic.
However, to the best of our knowledge, there have been no works designing uplink random access scheme for delay-constrained heterogeneous wireless networks.

In this paper, we take a first step to fill this blank by designing an RL-based random access scheme for a
delay-constrained heterogeneous wireless network with two users. 
Whilst one user (called user 1) adopts the slotted ALOHA scheme, we optimize the random access scheme of 
the other user (called user 2) with the goal of maximizing the system timely throughput.
We assume a distributed random access setting where user 2needs to design its scheme without knowing user 1's information. This is the most intriguing part of our problem.
Our major contributions of this paper are summarized as follows: 
\begin{itemize}
\item We first establish a  model-based  upper bound by assuming that user 2 has certain priori information of user 1.
For performance benchmarking, we derive a closed-form upper bound for the special case of hard deadline $D=1$
and derive a numerical upper bound based on an MDP formulation for general $D$.
\item We next propose an average-reward model-free RL-based random access scheme using R-learning \cite{schwartz1993reinforcement,singh1994reinforcement,sutton2018reinforcement}.
We illustrate that R-learning is more suitable than the widely-used discounted-award-based Q-learning for our problem,
 since the major performance metric, i.e., timely throughput, is
an average reward by nature. Since the state space of R-learning exponentially increases with $D$,
we further exploit the problem structure and design a low-complexity scheme by only utilizing the information about
whether user 2 has a most urgent packet (which will expire in one slot).
We call the proposed scheme \textbf{T}iny \textbf{S}tate-space \textbf{R}-learning random \textbf{A}ccess (TSRA).
\item Finally, we conduct extensive simulations and show that the system timely throughput of TSRA is $5.62\%$ higher than that of the existing baseline
DLMA \cite{yiding2019deep} and is only 4.98\% lower than the derived upper bound.
Furthermore, the time and space complexity of TSRA are respectively 80x and 17x reduced as compared with DLMA.
We also demonstrate the robustness of TSRA by considering different system settings.
\end{itemize}

%% file: model.tex
\section{System Model and Problem Formulation}\label{sec:system_model}
\begin{figure}[t]
  \centering
  \includegraphics[width=0.95\linewidth]{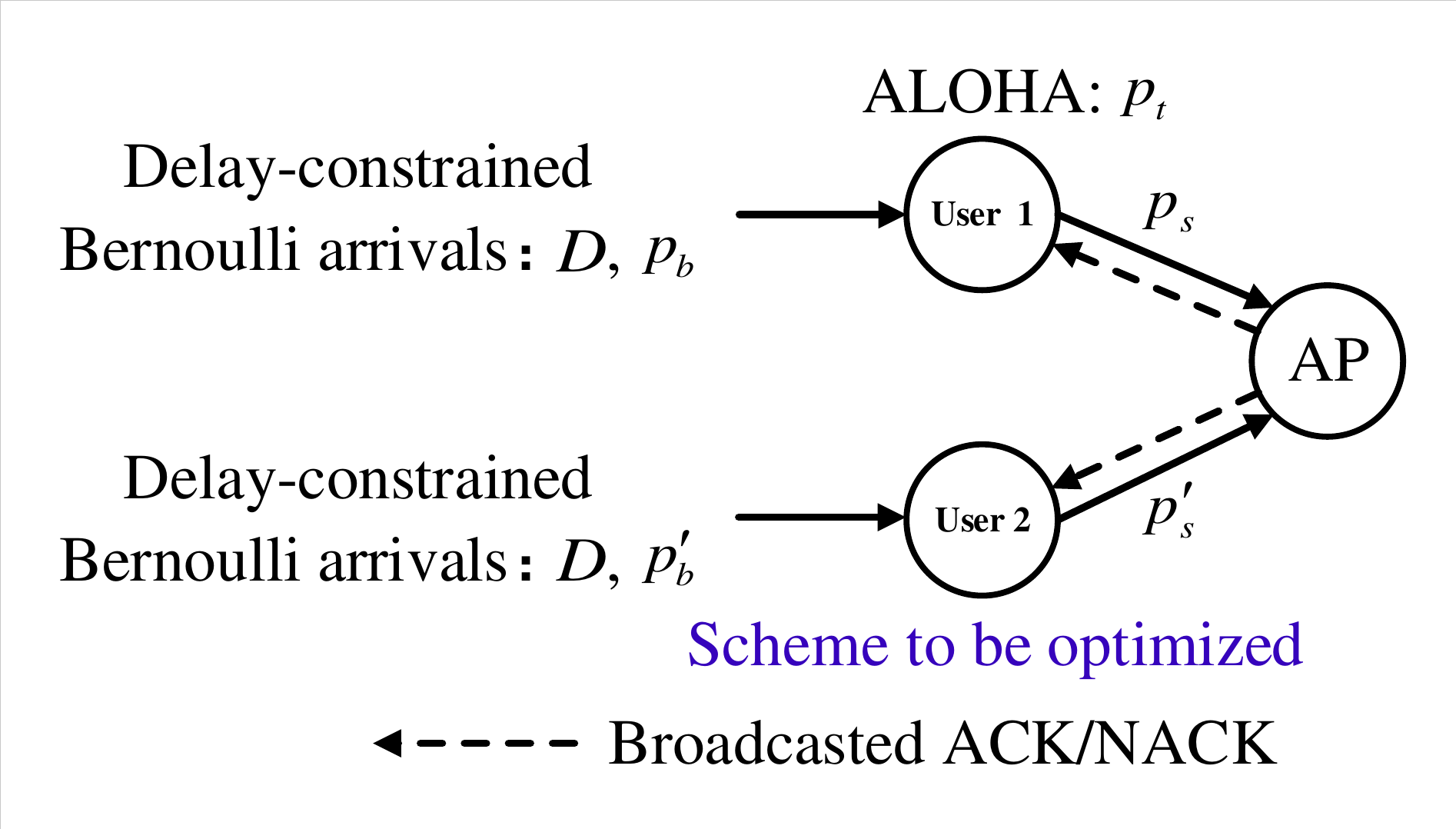}
  \caption{System model.}\label{fig:system_model}
\end{figure}

As a first step to study random access scheme for delay-constrained heterogeneous wireless networks,
we consider a two-user scenario in this paper, as shown in Fig.~\ref{fig:system_model}.
Specifically, two users share a wireless channel to
deliver delay-constrained traffic to an access point (AP). Time is slotted and indexed
from slot 1. We assume a delay-constrained Bernoulli traffic pattern for both users: user 1 (resp. user 2)
has a new packet arrival with  probability $p_b \in (0,1]$ (resp. $p'_b \in (0,1]$) in any slot, and
all packets have a hard delay of $D$ slots. A packet will be removed from the system
if it has not been delivered successfully to the AP in $D$ slots.

We assume an \emph{unreliable collision} wireless channel. If both users transmit a packet to the AP in a slot,
then a channel collision occurs and both packets cannot be successfully received by the AP. Even though only one user transmits a packet
to the AP, the wireless channel is still unreliable due to shadowing and fading. We model such unreliability by a
success probability. Specifically, if only user 1 (resp. user 2) transmits a packet to the AP, the packet
can be successfully delivered with probability $p_s \in (0,1]$ (resp. $p'_s \in (0,1]$). Otherwise,  a channel
error happens. Thus, a transmission failure may occur either due to a channel collision or due to a channel error.

The two-user network is heterogeneous in the sense that they use different random access schemes (i.e., transmission policies).
We assume that user 1 adopts the slotted ALOHA\footnote{For simplicity, we will use ALOHA to represent the slotted ALOHA in the rest of this paper.} protocol with transmission/retransmission probability $p_t \in [0,1]$.
Namely, user 1 always transmits or retransmits its head-of-line (HoL) packet to the AP with probability $p_t$ in any slot.
The random access scheme of user 2 is under our control. We design its random access scheme $\pi$
so as to maximize the system timely throughput \cite{lei2018on}, which is defined as
\bee
R^{\pi} \triangleq \liminf_{T \to \infty} \frac{\mathbb{E}^\pi \left[ \substack{\text{\# of packets of both users delivered successfully} \\ \text{ before expiration from slot $1$ to slot $T$}} \right]}{T}.
\label{equ:def-R}
\eee
The expectation is taken over all system randomness and possibly policy randomness.
Note that our design space is user 2's scheme while our goal is to maximize the system timely throughput.
This is in line with DARPA's vision on collaboratively-sharing scheme for spectrum  \cite{SC2,tilghman2019will}.

Note that both users cannot communicate with each other. Thus, user 2 does not know the queue information and transmission information of user 1.
However, it can interact with the environment (i.e., the AP) to learn the information of user 1.
Specifically, in the end of a slot, the AP will broadcast an acknowledgement (ACK) to both users if it successfully decodes a packet,
broadcast a negative-acknowledgement  (NACK) if it receives at least one packet but does not decode it successfully (either due to a channel collision or due to a channel error),
and broadcast nothing if it does not receive any packet in this slot.
By observing such feedback, user 2 aims to infer the behaviours of user 1
and then judiciously design its access scheme. This motivates us to use reinforcement learning (RL) \cite{sutton2018reinforcement}
due to its great success in solving many interactive decision problems in a model-free manner.

%% file: Upperbound.tex
\section{A Model-based Upper Bound} \label{sec:upper_bound}
Before we present our model-free RL-based random access scheme in Sec.~\ref{sec:TSRA},
we present a model-based upper bound in this section. Suppose
that user 2 is aware of user 1's parameters $p_b$, $D$, $p_t$, and $p_s$,
and queue information (i.e., the number of packets in the queue and the arrival time of each packet).
However, when user 2 decides to transmit or not transmit a packet at (the beginning of) any slot $t$,
it cannot know whether user 1 transmits/retransmits a packet or not at (the beginning of) slot $t$.
Otherwise, user 2 can always avoid collision such that the problem becomes trivial.
Such a model-based setting with more revealed information to user 2
allows us to derive an upper bound for the system timely throughput of the original model-free problem.
This upper bound will serve as a performance benchmark for evaluating any random access policy later.

We first consider the special case of hard deadline $D=1$. Note that
$D=1$ means that any packet arriving at (the beginning of) slot $t$
will expire at the end of slot $t$ if it is not transmitted or its transmission fails
due to channel collision or channel error. Thus, in any slot,
the queue of both users has at most one packet. This significantly simplifies
the system design due to coupling-free between different slots. We can thus
derive the optimal policy of user 2, which surprisingly is a binary policy, as shown in the following theorem.

\begin{theorem} \label{thm:D=1}
If $D=1$, then the optimal strategy of user 2 to maximize the system timely throughput is i) always transmitting the HoL packet if
\be
p_bp_t < \frac{p'_s}{p_s + p'_s}, \label{equ:always-transmit-D=1}
\ee
and ii) always remaining idle if its queue is empty or \eqref{equ:always-transmit-D=1} does not hold.
The corresponding system timely throughput is
\be
R=
\left\{
  \begin{array}{ll}
    \left[ p'_{s}-(p_{s}+p'_{s})p_{t}p_{b}\right]p'_{b}+p_{s}p_{t}p_{b}, & \hbox{if \eqref{equ:always-transmit-D=1} holds;} \\
    p_{s}p_{t}p_{b}, & \hbox{otherwise.}
  \end{array}
\right.
\label{equ:upper-bound-D=1}
\ee
\end{theorem}
\begin{IEEEproof}
Please refer to Appendix~\ref{app:proof-of-theorem-D=1}.
\end{IEEEproof}

Let us investigate condition \eqref{equ:always-transmit-D=1} by assuming that $p_s=p'_s$, i.e., both users have the same channel quality.
In this case, condition \eqref{equ:always-transmit-D=1} becomes $p_bp_t < 0.5$. Note that $p_b p_t$ is
the probability that user 1 transmits a packet in any slot since $D=1$. Thus, if this probability is less than 0.5, i.e., user 1 is not aggressive,
user 2 will become completely aggressive to take over the channel.
On the contrary, if this probability is larger than 0.5, i.e., user 1 is aggressive,
user 2 will become completely unaggressive to hand over the channel.
Such a binary policy achieves the best tradeoff between utilizing the wireless channel
and avoiding collision. The closed-form expression \eqref{equ:upper-bound-D=1} serves as an upper bound for $D=1$.

However, for general $D$, it is difficult to directly characterize an optimal strategy and the corresponding system timely throughput
because of the coupling between different slots. We thus formulate our problem as an MDP problem
and propose an upper bound by solving a linear programming problem. An MDP is characterized by its state, action, reward function, and transition probability \cite{puterman1994markov}.
The state of the system at (the beginning of) slot $t$ is defined as
\be
s_t \triangleq (l_{t,1}, l_{t,2}, o_t). \label{equ:S1}
\ee
In \eqref{equ:S1}, $l_{t,i}=(l^1_{t,i},l^2_{t,i},\cdots,l^D_{t,i})$ is the lead time vector \cite{lei2017timely} of user $i \in \{1,2\}$ at slot $t$, where
\be
l^k_{t,i} =
\left\{
  \begin{array}{ll}
    1, & \hbox{if user $i$ has a packet at slot $t$, which} \\
       & \hbox{will expire in $k \in \{1,2,\cdots,D\}$ slots;} \\
    0, & \hbox{otherwise.}
  \end{array}
\right.
\nnb
\ee
Further, $o_t \in \{\textsf{BUSY}, \textsf{SUCCESSFUL}, \textsf{IDLE}, \textsf{FAILED}\}$ is the channel observation at (the beginning of) slot $t$,
equivalently, at the end of slot $t-1$. Specifically, channel observation $o_t=\textsf{BUSY}$
means that user 2 does not transmit a packet but receives an ACK from the AP in slot $t-1$,
indicating that user 1 transmits a packet and no channel error happens in slot $t-1$.
Channel observation $o_t=\textsf{SUCCESSFUL}$ means that user 2 transmits a packet
and receives an ACK from the AP in slot $t-1$, indicating that
user 1 does not transmit a packet and there is no channel error for user 2's packet.
Channel observation $o_t=\textsf{IDLE}$ means that user 2 receives nothing from the AP at the end of slot $t-1$,
indicating that both users do not transmit a packet in slot $t-1$.
Channel observation $o_t=\textsf{FAILED}$ means that user 2 receives a NACK from the AP at the end of slot $t-1$,
indicating that a channel collision or a channel error happens. Without loss of generality, we assume that $o_1=\textsf{IDLE}$.
We remark that the modeling for channel observation  is the same as \cite{yiding2019deep}.
The set of all possible states is denoted by $\mathcal{S}$.
Clearly, we have $\left|\mathcal{S} \right|=2^D \cdot 2^D \cdot 4 = 2^{2D+2}$.

At slot $t$, the action of user 2 is denoted by $a_t$. Similar to \cite{yiding2019deep}, the action space is defined as $\mathcal{A} \triangleq \{\textsf{TRANSMIT}, \textsf{WAIT}\}$.
One can readily prove that it is optimal to first transmit the HoL packet if there are multiple packets in the user 2's queue at any slot.
Thus, action $a_t=\textsf{TRANSMIT}$ means that user 2 transmits its HoL packet at slot $t$, while $a_t=\textsf{WAIT}$ means that
user 2 does not transmit a packet at slot $t$.

We define the reward function $r(s_t,a_t)$ as
\bee
r(s_t,a_t) \triangleq 1_{\left\{o_t \in \{ \textsf{BUSY}, \textsf{SUCCESSFUL}\} \right \}}, \forall s_t \in \mathcal{S}, o_t \in \mathcal{A},
\label{equ:reward function}
\eee
where $1_{\{\cdot\}}$ is the indicator function. Note that $o_t=\textsf{BUSY}$
means that user 1 transmits a packet successfully in slot $t-1$,
and $o_t= \textsf{SUCCESSFUL}$ means that user 2 transmits a packet successfully in slot $t-1$.
Thus $r(s_t,a_t)=1$ if the system (either user 1 or user 2) transmits a packet successfully in slot $t-1$.
Note that we model the reward with ``delay of gratification". Namely,
the delivered packet in slot $t-1$ is translated into the reward at slot $t$.
However, since our performance metric is long-term system timely throughput,
such ``delay of gratification" will not cause performance loss, as shown in \eqref{equ:R-is-same-as-avg-reward} later.
In addition, we remark that the reward function only depends on the channel observation $o_t$, regardless of system state $s_t$.

The transition probability from state $s$ to state $s'$ if taking action $a$ is defined as
\be
P(s'|s,a) \triangleq P(s_{t+1}=s'|s_{t}=s, a_t=a), \forall t, s,s',a, \label{equ:MDP-P}
\ee
which depends on
(i) the arrival and expiration events of both users,
(ii) the transmission events of both users,
(iii) the channel collision and channel error events,
and (iv) the change of lead time vector.
We use an example to illustrate how to compute the transition probabilities; see \texttt{https://github.com/DanzhouWu}
\texttt{/TSRA/tree/main/TransitionProbaility}.

Based on the above MDP model, it is straightforward to see that the system timely throughput under a policy $\pi$ defined in \eqref{equ:def-R} is equivalent
to the average reward of our formulated MDP under policy $\pi$, i.e.,
\bee
R^{\pi} & = \liminf_{T \to \infty} \frac{\mathbb{E}^\pi \left[ \substack{\text{\# of packets of both users delivered successfully} \\ \text{ before expiration from slot $1$ to slot $T$}} \right]}{T}, \nnb \\
 & =  \liminf_{{T}\rightarrow\infty} \frac{\sum_{t=2}^{T+1} \mathbb{E}^{\pi}\{r(s_t, a_t)\} }{{T}}, \nnb \\
 & =  \liminf_{{T}\rightarrow\infty} \frac{\sum_{t=1}^{T} \mathbb{E}^{\pi}\{r(s_t, a_t)\} }{{T}}. \label{equ:R-is-same-as-avg-reward}
\eee

Thus, our problem becomes an average-reward MDP problem.
Here we use the dual linear program approach to solve this MDP problem \cite[Chapter 9.3]{puterman1994markov},
\bee
\max & \quad \sum_{s \in \mathcal{S}} \sum_{a \in \mathcal{A}} r(s, a) x(s,a) \nnb \\
\text{s.t.} & \quad \sum_{a \in \mathcal{A}} x(s',a) = \sum_{s \in \mathcal{S}} \sum_{a \in \mathcal{A}} P(s'|s, a) x(s,a), \quad \forall s' \in \mathcal{S} \nnb \\
& \quad \sum_{a \in \mathcal{A}} x(s',a) + \sum_{a \in \mathcal{A}} y(s',a)\nnb \\
& \quad \quad= \sum_{s \in \mathcal{S}} \sum_{a \in \mathcal{A}} P(s'|s, a) y(s, a) + \alpha_{s'}, \quad \forall s' \in \mathcal{S} \quad \nnb \\
\text{var.} & \quad x(s, a) \geq 0, \quad y(s, a) \geq 0, \label{equ:MDP-LP}
\eee
where $\{\alpha_{s}: s \in \mathcal{S}\}$ are arbitrary constants such that $\alpha_s > 0 \;\; (\forall s \in \mathcal{S})$ and $\sum_{s \in\mathcal{S}}\alpha_{s} = 1$. Note that we follow standard procedures of the dual linear program
 in \cite[Chapter 9.3]{puterman1994markov}. Basically, notation
$x(s,a)$ (resp. $y(s,a)$) represents the frequency (or the stationary probability) that
the Markov chain is on state $s$ and the action is $a$ where $s$ is a recurrent state (resp. a transient state);
see \cite[Proposition 9.3.2]{puterman1994markov}.

The optimal value of problem \eqref{equ:MDP-LP} serves as an upper bound for general $D$. In addition, according to \cite{hordijk1979linear} and \cite[Chapter 9.3.1]{puterman1994markov},
solving problem \eqref{equ:MDP-LP} also yields a randomized optimal policy,
\bee
\pi(a|s) =
\left\{
  \begin{array}{ll}
    \frac{x^*(s,a)}{\sum_{a\in \mathcal{A}} x^*(s,a)}, & \hbox{if $\sum_{a\in \mathcal{A}}{x^*(s,a)} > 0$;} \\
    \frac{y^*(s,a)}{\sum_{a\in \mathcal{A}} y^*(s,a)}, & \hbox{otherwise.}
  \end{array}
\right.
\label{equ:upper-bound-policy}
\eee
where $\pi(a|s)$ is the probability of taking action $a$ under state $s$,
and $\{x^*(s,a), y^*(s,a): s \in \mathcal{S}, a \in \mathcal{A}\}$ is an optimal solution of problem \eqref{equ:MDP-LP}.

%% file: TSRA.tex
\section{Tiny State-space R-learning Random Access }\label{sec:TSRA}

The disadvantage of model-based MDP is that user 2 needs to know user 1's parameters and queue information. However, these information cannot
be obtained in practice such that user 2 cannot know user 1's queue state $l_{t,1}$ and
the transition probabilities $P(s'|s,a)$ (Please refer to \eqref{equ:S1} and \eqref{equ:MDP-P}).
To address this issue, reinforcement learning (RL) has been proposed as a model-free approach to solve MDP problems.
RL needs the state space $\mathcal{S}$, the action space $\mathcal{A}$, and the reward function $r(s,a), \forall s \in \mathcal{S}, a \in \mathcal{A}$,
but does not need the transition probabilities $P(s'|s,a)$ of an MDP. Instead, RL learns the model by directly interacting with the environment.

Since user 2 cannot know user 1's queue information, we define its state at slot $t$ as\footnote{With a little bit abuse of notation,
in the model-free problem in this section, except for the state space, we adopt the same notations of the model-based problem in Sec.~\ref{sec:upper_bound}.
Namely, we still use $s_t$ to denote
the state, $a_t$ to denote the action, and $r(s_t,a_t)$ to denote the reward function for the model-free problem in this section.
They are distinguishable in the context.},
\be
s_t \triangleq (l_{t,2}, o_t), \label{equ:equ-state-FSQA-and-FSRA}
\ee
where $l_{t,2}$ is the queue information of user 2 itself, and $o_t$ is the channel observation (same as Sec.~\ref{sec:upper_bound}).
The state space $\mathcal{S}'$ is thus of size $2^D \cdot 4 = 2^{D+2}$.
The action space  $\mathcal{A} = \{\textsf{TRANSMIT}, \textsf{WAIT}\}$ is again the same as Sec.~\ref{sec:upper_bound}.
The reward function $r(s_t,a_t)$ is defined as
\bee
r(s_t,a_t) \triangleq 1_{\left\{o_t \in \{ \textsf{BUSY}, \textsf{SUCCESSFUL}\} \right \}}, \forall s_t \in \mathcal{S}', a_t \in \mathcal{A},
\label{equ:reward function_RL}
\eee
which is similar to that in the model-based setting (Please refer to \eqref{equ:reward function}).
Namely, the reward is 1 if user 2 receives an ACK, either for its own packet ($o_t=\textsf{SUCCESSFUL}$) or for user 1's packet ($o_t=\textsf{BUSY}$).

\subsection{Q-Learning} \label{subsec:Q-learning}
Based on the above information, we can apply different RL methods to solve our problem in a model-free manner,
such as Monte Carlo, temporal-difference learning, etc. \cite{sutton2018reinforcement}.
Among them, Q-learning is one of the most widely-used methods \cite{sutton2018reinforcement}.
In fact, the delay-unconstrained counterpart of our problem, i.e., \cite{yiding2019deep}, also used Q-learning.
The simplest form of Q-learning, called one-step Q-learning, iteratively updates the Q-function $Q(s,a)$ as follows,
\bee
Q(s_t, a_t) & \leftarrow Q (s_t, a_t) + \alpha \big[r(s_t, a_t) + \nnb \\
& \qquad \gamma \max_{a}Q(s_{t+1}, a) - Q(s_t, a_t) \big], \label{equ:Q-learning-Q-function}
\eee
where $\alpha \in (0,1]$ is the learning rate, $\gamma \in (0,1)$ is the discount factor, and
$Q(s,a)$ is the state-action value function (called Q-function),  approximating the discounted reward for given state and action
for the iteratively updated policy $\pi$, i.e.,
\be
Q(s,a) \approx \mathbb{E}^{\pi} \left[\sum_{\tau=t}^{\infty} \gamma^{\tau-t}r(s_\tau, a_\tau) | s_t=s, a_t=a \right]. \label{equ:Q-function-approx}
\ee
Note that the policy $\pi$ is iteratively updated by selecting action $a$ to maximize $Q(s,a)$ for any state $s$
with an $\epsilon$-greedy algorithm \cite{sutton2018reinforcement}.
We call the algorithm \textbf{F}ull \textbf{S}tate-space \textbf{Q}-learning random \textbf{A}ccess (FSQA),
which is detailed in Algorithm~\ref{alg:ql}.

Q-learning is suitable for solving MDPs with discounted reward in a model-free manner.
However, in network communication research, the major performance metric, throughput or timely throughput, is a long-term average reward.
Therefore, Q-learning may be less suitable for network communication research than another RL method, called R-learning,
which solves MDPs with average reward in a model-free manner \cite{schwartz1993reinforcement, singh1994reinforcement,sutton2018reinforcement}.

\begin{algorithm}[t]
 \caption{FSQA Algorithm For User 2}
 \label{alg:ql}
\begin{algorithmic}[1]
  \STATE Initialize Q-function $Q(s, a) = 0$, $\forall s \in \mathcal{S}'$, $\forall a \in \mathcal{A}$,
    \STATE Set learning rate $\alpha=0.01$
    \STATE Initialize the discount factor $\gamma=0.9$
    \STATE Observe the initial system state $s_1$
    \FOR{$t = 1, 2, \cdots$}
    \STATE Choose $a_t$ with an $\epsilon$-greedy algorithm, i.e.,
    \[
    a_t=
    \left\{
      \begin{array}{ll}
        \arg \max_{a} Q(s_t, a), & \hbox{with prob. $1-\epsilon_t$;} \\
        \text{random action}, & \hbox{with prob. $\epsilon_t$,}
      \end{array}
    \right.
    \]
    where $\epsilon_t=\max\{0.995^{t-1},0.01\}$
    \STATE Observe $r(s_t, a_t)$, $s_{t+1}$
    \STATE Update Q-function as follows,
    		\bee
	           Q(s_t, a_t) & \leftarrow Q(s_t, a_t) + \alpha \big( r(s_t, a_t) \nnb \\
						&\quad + \gamma \max_{a}Q(s_{t+1}, a) - Q(s_t, a_t) \big) \nnb
			\eee
    \ENDFOR
\end{algorithmic}
\end{algorithm}

\subsection{R-Learning} \label{subsec:R-learning}

R-learning also utilizes the state-action value function (we still call it Q-function by convention), which however has a different meaning.
Among the variants of R-learning \cite{schwartz1993reinforcement, singh1994reinforcement,sutton2018reinforcement},
in this paper, we adopt the version in \cite[Algorithm 3]{singh1994reinforcement} and \cite[Figure 11.2]{sutton2018reinforcement},
\bee
Q(s_t, a_t) & \leftarrow Q(s_t, a_t) + \alpha \big[ r(s_t, a_t) + \nnb \\
& \qquad \max_{a}Q(s_{t+1}, a) - Q(s_t, a_t) - \rho \big], \label{equ:upgrade_Q}\\
& \rho \leftarrow \rho + \beta\big[ r(s_t, a_t) + \nnb \\
& \qquad \max_{a}Q(s_{t+1}, a) - Q(s_t, a_t) - \rho \big], \label{equ:upgrade_rho}
\eee
where $\alpha \in (0,1]$ and $\beta \in (0,1]$ are learning rates, $\rho$ approximates the state-independent
average reward for the iteratively updated policy $\pi$, i.e.,
\be
\rho \approx \lim_{T \to \infty} \mathbb{E}^{\pi} \left[ \frac{\sum_{t=1}^T r(s_t,a_t)}{T} \right],
\label{equ:rho-approx-R-learning}
\ee
and Q-function $Q(s, a)$ approximates the state-dependent cumulative reward difference (called relative value in \cite{singh1994reinforcement,sutton2018reinforcement}) for the iteratively updated policy $\pi$, i.e.,
\be
Q(s,a) \approx  \mathbb{E}^{\pi}  \left[ \left. \sum_{\tau=t}^{\infty}  \left[r(s_\tau,a_\tau) - \rho \right] \right| s_t = s, a_t=a \right].
\label{equ:Q-approx-R-learning}
\ee
Similar to Q-learning, the policy $\pi$ is iteratively updated by selecting action $a$ to maximize $Q(s,a)$ for any state $s$
with an $\epsilon$-greedy algorithm.
We call the algorithm \textbf{F}ull \textbf{S}tate-space \textbf{R}-learning random \textbf{A}ccess (FSRA),
which is detailed in Algorithm~\ref{alg:rl}.

\begin{algorithm}[h]
 \caption{FSRA/HSRA/TSRA Algorithm for User 2}
 \label{alg:rl}
\begin{algorithmic}[1]
    \STATE Initialize Q-function $Q(s, a) = 0, \forall a \in \mathcal{A}$, $\forall s \in \tilde{\mathcal{S}}$, where the state space $\tilde{\mathcal{S}}$
is different for different algorithms,
    \[
    \tilde{\mathcal{S}}=
    \left\{
      \begin{array}{ll}
         \mathcal{S}', & \hbox{If the algorithm is FSRA;} \\
         \mathcal{S}'', & \hbox{If the algorithm is HSRA;} \\
         \mathcal{S}''', & \hbox{If the algorithm is TSRA;} \\
      \end{array}
    \right.
    \]
    \label{line:diff-state-space}
    \STATE Initialize $\rho = 0$
    \STATE Set learning rates $\alpha=0.01$, $\beta=0.01$
    \STATE Observe the initial system state $s_1$
    \FOR{$t = 1, 2, \cdots$}
    \STATE Choose $a_t$ with an $\epsilon$-greedy algorithm, i.e.,
    \[
    a_t=
    \left\{
      \begin{array}{ll}
        \arg \max_{a} Q(s_t, a), & \hbox{with prob. $1-\epsilon_t$;} \\
        \text{random action}, & \hbox{with prob. $\epsilon_t$,}
      \end{array}
    \right.
    \]
    where $\epsilon_t=\max\{0.995^{t-1},0.01\}$
    \STATE Observe $r(s_t, a_t)$, $s_{t+1}$
    \STATE Update Q-function as follows,
    		\bee
			Q(s_t, a_t) & \leftarrow Q(s_t, a_t) + \alpha \big( r(s_t, a_t)  + \nnb \\
						&\qquad \max_{a}Q(s_{t+1}, a) - Q(s_t, a_t) - \rho \big) \nnb
			\eee
    \STATE Update $\rho$ as follows,
    		\bee
    		\rho & \leftarrow \rho + \beta\big( r(s_t, a_t) +  \nnb \\
    				&\qquad \max_{a}Q(s_{t+1}, a) - Q(s_t, a_t) - \rho \big) \nnb
    		\eee
    \ENDFOR
\end{algorithmic}
\end{algorithm}

\begin{figure}[t]
  \centering
  \includegraphics[width=0.9\linewidth]{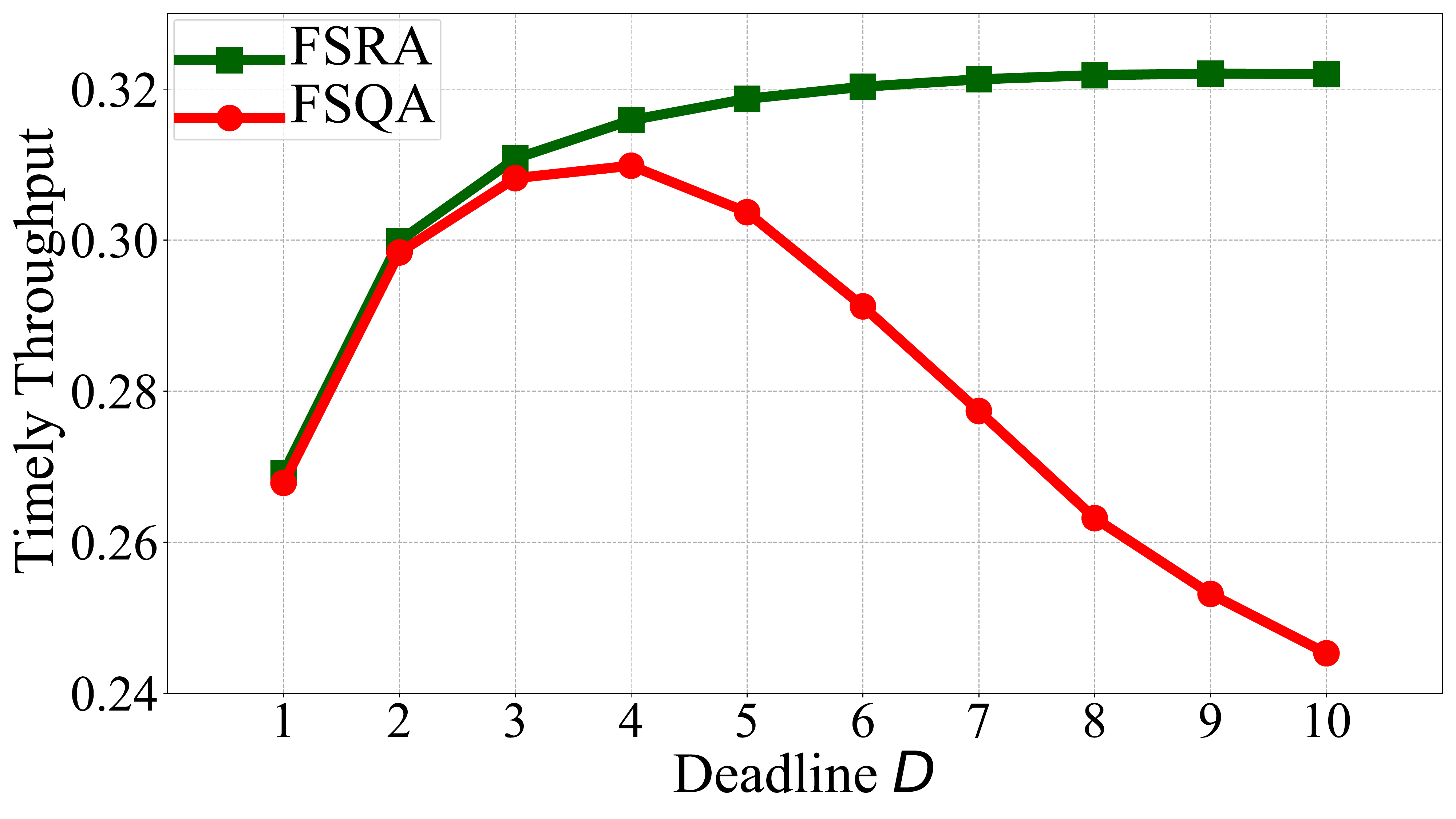}
  \caption{Comparison of the system timely throughputs of FSRA and FSQA.}\label{fig:compare_rl_ql}
\end{figure}

We compare FSQA and FSRA by ranging $D$ from 1 to 10.
For each $D$, we randomly select 500 groups of different system parameters, i.e., ($p_b$, $p_b'$, $p_s$, $p_s'$, $p_t$).
For each group of parameters, we simulate 10,000,000 slots independently for FSQA and FSRA,
and evaluate the system timely throughput for the last 100,000 slots. The result is shown in Fig.~\ref{fig:compare_rl_ql}.
As we can see, FSRA outperforms FSQA for all $D$'s, suggesting that indeed R-learning is more suitable to our problem than Q-learning.
As we explained before, R-learning is used to solve model-free MDPs with average reward, while
Q-learning is used to solve model-free MDPs with discounted reward. Our problem turns out to be exactly
a model-free MDP with average reward. That is the main reason that FSRA outperforms FSQA.
In Appendix~\ref{app:how_to_improve_FSQA}, we further present an example to compare the policies of FSRA and FSQA after convergence
and explicitly show that FSRA is better than FSQA. We also propose a method to tune FSQA so as to improve its performance.

In addition, we remark that FSQA is more difficult to converge than FSRA. As the deadline $D$ increases,
we should expect that the system timely throughput also increases since packets have longer lifetime and thus are more difficult to expire.
However, as $D$ increases, the state space $\mathcal{S}'$ (of size $2^{D+2}$) also increases exponentially.
As a result, both FSRA and FSQA needs more slots to converge. But FSQA is much more sensitive to the state-space explosion.
When $D \ge 4$, FSQA cannot converge in 10,000,0000 slots such that its system timely throughput even decreases as $D$ increases, as shown in Fig.~\ref{fig:compare_rl_ql}. We will explicitly compare the convergence speeds of FSQA and FSRA in Sec.~\ref{subsec:compare-convergence}.

Although FSRA outperforms FSQA in terms of both the achieved system timely throughput and the convergence speed,
we point out that FSRA still converges slowly as $D$ increases. For example, in Fig.~\ref{fig:compare_rl_ql},
we need to run 10,000,000 slots such that FSRA converges.
This problem is even more severe when $D$ is larger since
the state space $\mathcal{S}'$ (of size $2^{D+2}$) increases exponentially with deadline $D$. Please refer to Sec.~\ref{subsec:compare-convergence}
to see the slow convergence speed of FSRA.
This disadvantage is not acceptable for highly dynamic heterogeneous wireless networks, since a small change
of the network could cause the system to take a long time to re-converge.
To address this problem, we further explore the problem structure and significantly reduce the state space.
As we mentioned in Sec.~\ref{sec:upper_bound},
it is optimal to first transmit the HoL packet (the most urgent packet) if there are multiple packets in the user 2's queue at any slot.
Thus, we can imagine that the HoL packet has the biggest impact on the system performance.
In fact, \cite{zhang2020scheduling} has applied this idea to derive a near-optimal
heuristic scheduling policy only based on the lead time of the HoL packet
for wireless downlink with deadline and retransmission constraints.
We can also design a new R-learning random access algorithm only based on the lead time of the HoL packet.
Namely, the state of user 2 at slot $t$ becomes
\be
s_t \triangleq (h_{t,2}, o_t), \label{equ:state-HSRA}
\ee
where $h_{t,2}$ is the lead time of the HoL packet of user 2 at slot $t$ and it is 0 by convention
if user 2 does not have any packet at slot $t$. The state space is denoted by $\mathcal{S}''$,
which is of size $4(D+1)$. The R-learning based algorithm is the same as FSRA except that
the state space changes from $\mathcal{S}'$ to $\mathcal{S}''$. We call this algorithm
\textbf{H}oL-packet-based \textbf{S}tate-space \textbf{R}-learning random \textbf{A}ccess (HSRA),
which is detailed in Algorithm~\ref{alg:rl}.

We can be even more aggressive by only considering if user 2 has a packet whose lead time is 1.
A packet with lead time 1 means that it will be expire at the end of the current slot if it cannot be delivered
successfully in the current slot. Thus, such a packet is the most urgent one among all packets
in the system. Therefore, we re-define the system state of user 2 as
\be
s_t \triangleq (f_{t,2}, o_t), \label{equ:state-TSRA}
\ee
where
\be
f_{t,2} =
\left\{
  \begin{array}{ll}
    1, & \hbox{if user 2 has a packet whose lead time is 1;} \\
    0, & \hbox{otherwise.}
  \end{array}
\right.
\label{equ:def-f-t-2}
\ee
The state space is denoted by $\mathcal{S}'''$ whose size is only 8 now.
Since the state space is quite small and even not related to deadline $D$,
we call this algorithm \textbf{T}iny \textbf{S}tate-space \textbf{R}-learning Random \textbf{A}ccess (TSRA).
Again, TSRA is the same as FSRA except that the state space changes from $\mathcal{S}'$ to $\mathcal{S}'''$,
which is also detailed in Algorithm~\ref{alg:rl}.
Since the state space of TSRA is quite small, it converges much faster than FSRA, as shown in Sec.~\ref{subsec:compare-convergence} shortly.
We will also show that its performance is close to HSRA and FSRA in Sec.~\ref{sec:simulation}.
Thus, this is the final designed policy for our studied problem in Sec.~\ref{sec:system_model}.

\begin{figure}[t]
  \centering
  \includegraphics[width=0.9\linewidth]{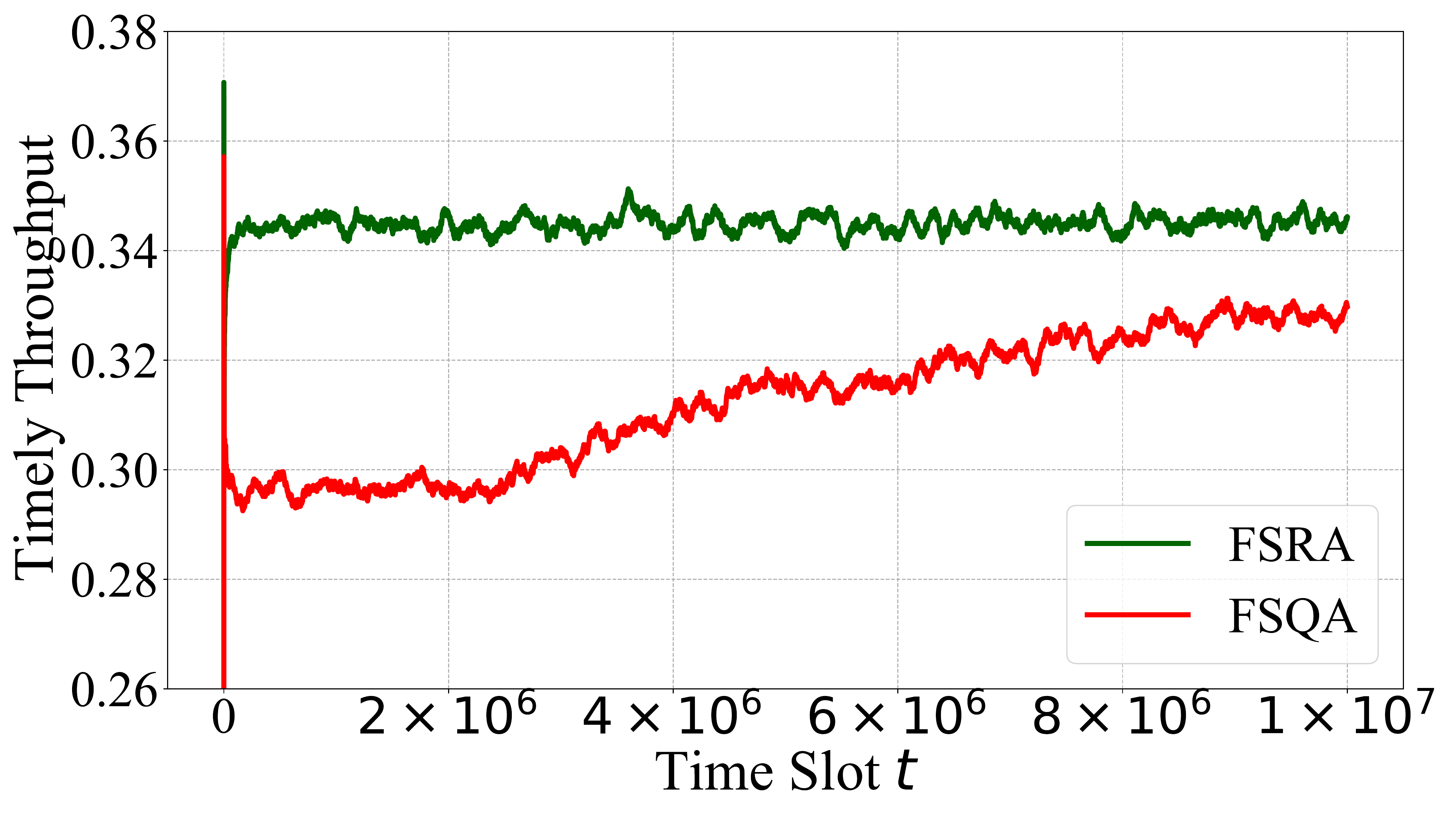}
  \caption{Comparison of the convergence speeds of FSRA and FSQA where $p_b=0.5$, $p_b'=0.4$, $p_s=0.7$, $p_s'=0.6$, $p_t=0.4$, $D=5$.}\label{fig:convergence_fsqa}
\end{figure}

\subsection{Comparing Convergence Speeds of FSQA, FSRA and TSRA} \label{subsec:compare-convergence}
In this subsection, we compare the convergence speeds of FSQA, FSRA and TSRA.

We first show that FSQA is more difficult to converge than FSRA.  We set the system parameters $p_b=0.5$, $p_b'=0.4$, $p_s=0.7$, $p_s'=0.6$, $p_t=0.4$, $D=5$.
The result is shown in Fig.~\ref{fig:convergence_fsqa}.
We can observe that FSRA converges in 200,000  slots, while FSQA does not converge at the end of the simulation.
Namely, FSQA cannot converge in 10,000,000 slots in this example, which indeed demonstrates that FSQA is very difficult to converge.

\begin{figure}[t]
  \centering
  \includegraphics[width=0.9\linewidth]{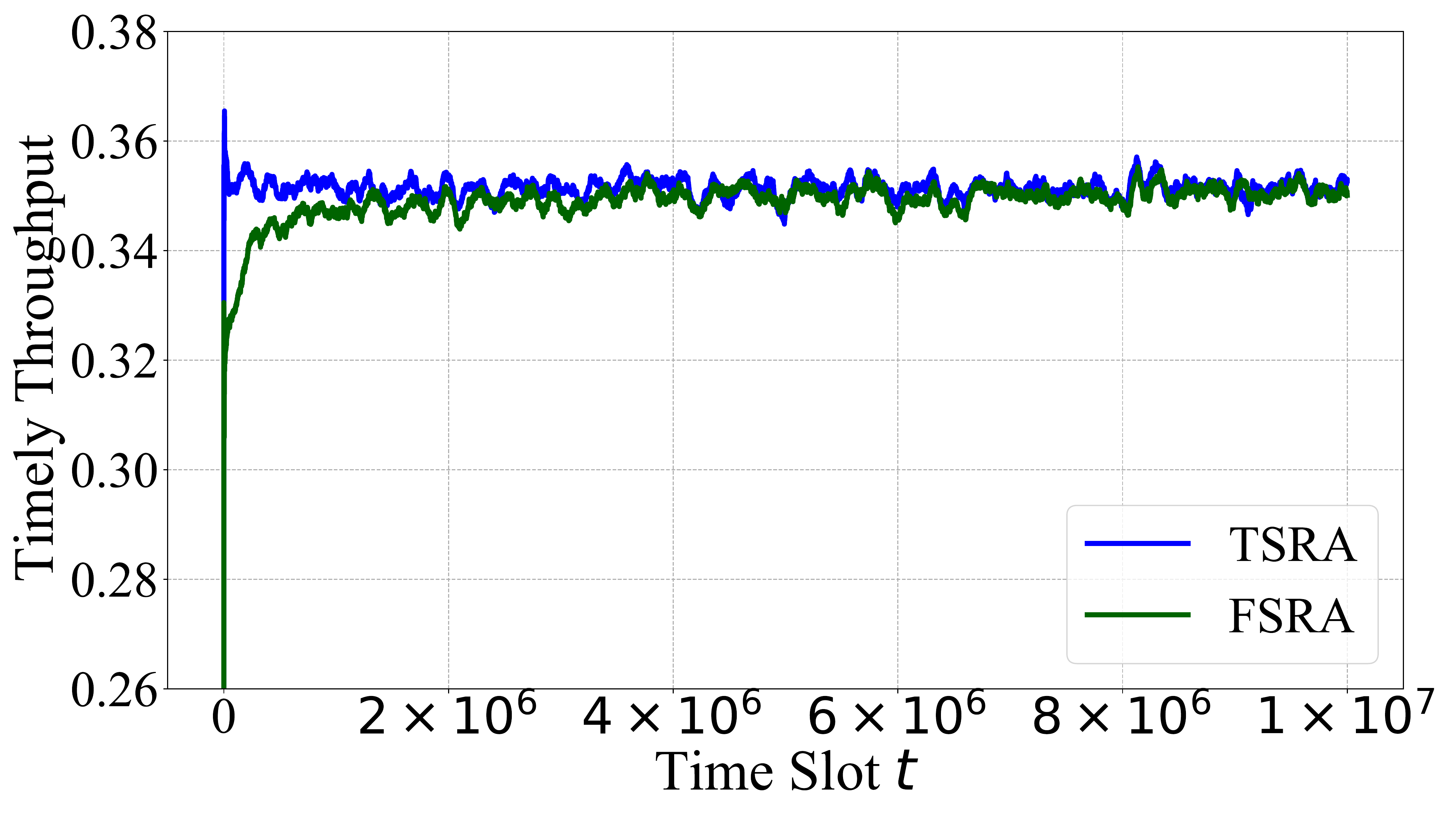}
  \caption{Comparison of the convergence speeds of TSRA and FSRA where $p_b=0.5$, $p_b'=0.4$, $p_s=0.7$, $p_s'=0.6$, $p_t=0.4$, $D=10$.}\label{fig:convergence_fsra}
\end{figure}

We next show that FSRA is more difficult to converge than TSRA for large $D$.
We again set the system parameters $p_b=0.5$, $p_b'=0.4$, $p_s=0.7$, $p_s'=0.6$, $p_t=0.4$, but with a large deadline $D=10$.
The result is shown in Fig.~\ref{fig:convergence_fsra}.
We can observe that TSRA converges much faster than FSRA, and its achieved system
timely throughput after convergence is almost the same as that of FSRA.

\begin{figure}[t]
  \centering
  \includegraphics[width=0.9\linewidth]{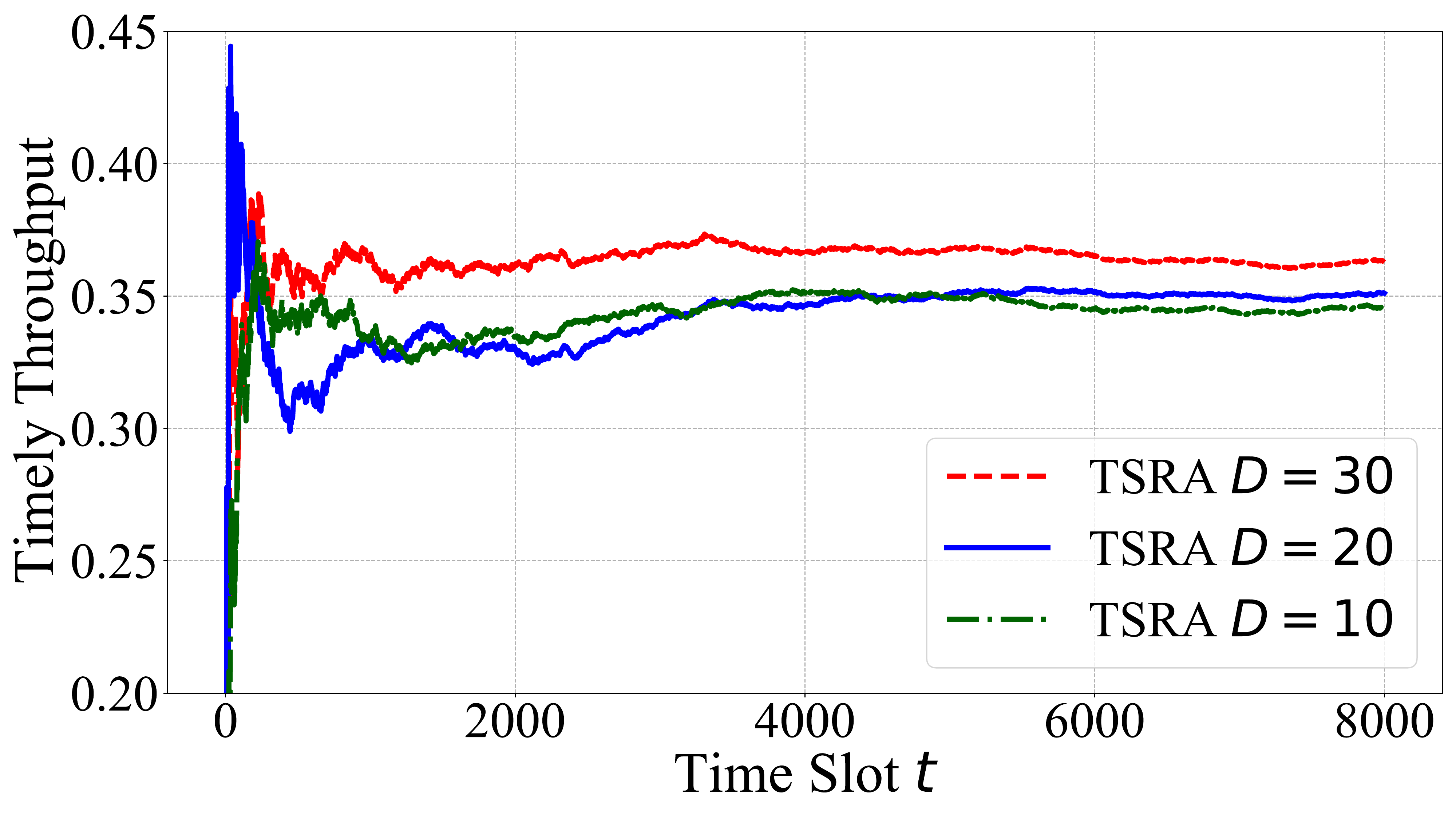}
  \caption{Comparison of the convergence speeds of TSRA where $p_b=0.5$, $p_b'=0.4$, $p_s=0.7$, $p_s'=0.6$, $p_t=0.4$.}\label{fig:convergence_tsra}
\end{figure}

We can further enlarge deadline $D$ and show that TSRA still converges very fast.
We again set the system parameters $p_b=0.5$, $p_b'=0.4$, $p_s=0.7$, $p_s'=0.6$, $p_t=0.4$, and let $D$ be 10, 20, and 30, respectively.
The results are shown in Fig.~\ref{fig:convergence_tsra}. We can observe that TSRA converges in 6,000 time slots for all three cases.
The fast convergence speed of TSRA makes it suitable in practical highly-dynamic heterogeneous wireless networks.

%% file: simulation.tex
\section{Simulations}\label{sec:simulation}

\begin{figure}[t]
  \centering
  \includegraphics[width=0.9\linewidth]{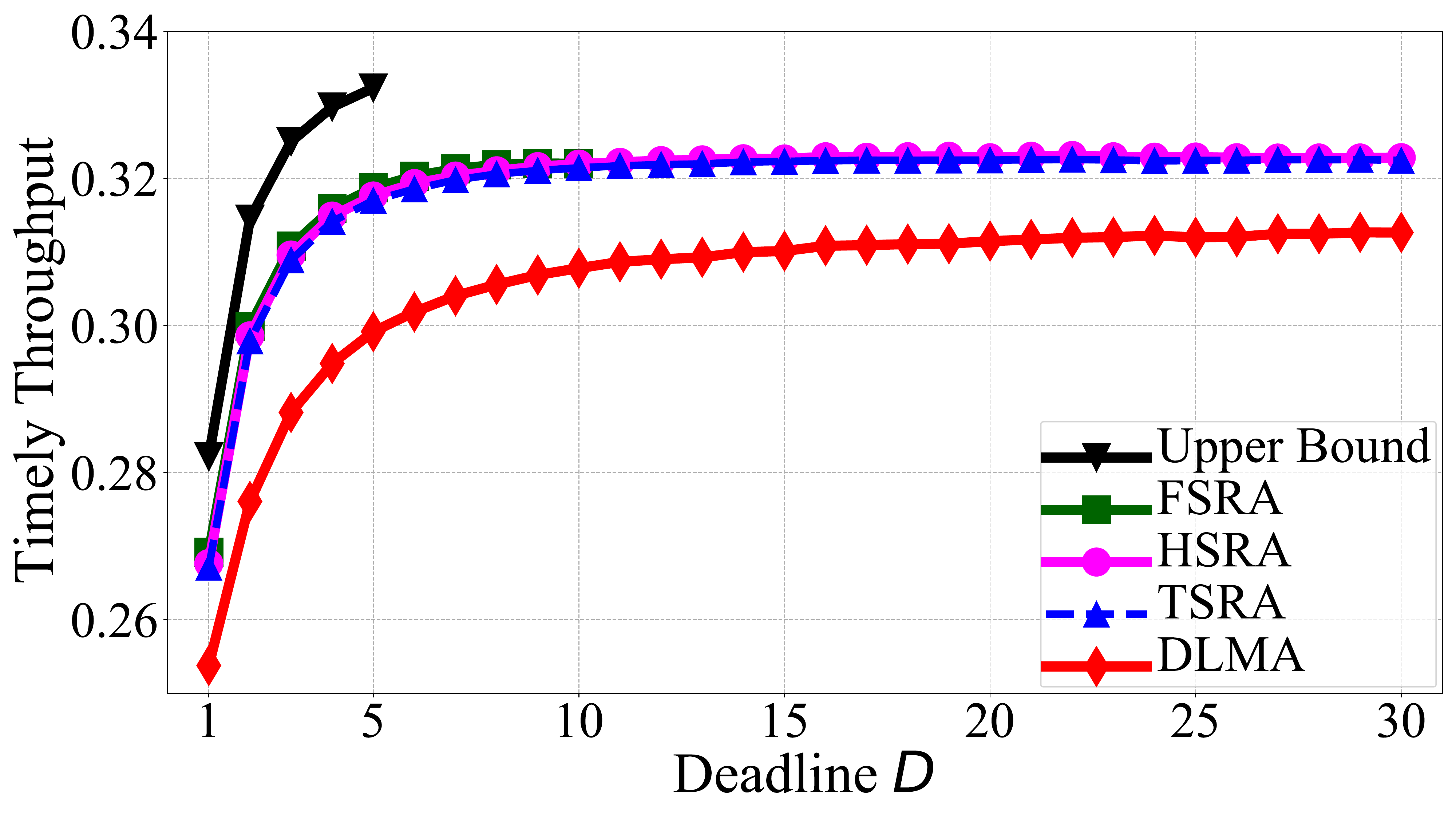}\\
  \caption{Comparison of the system timely throughputs of the upper-bound algorithm, FSRA/HSRA/TSRA, and DLMA. \label{subfig:fig:timely_throu}}
\end{figure}

In this section, we carry out extensive simulations to validate the effectiveness
of our proposed random access scheme TSRA and demonstrate that TSRA outperforms the existing baseline,
DLMA, which is the random access scheme adopted by \cite{yiding2019deep} for delay-unconstrained heterogeneous wireless networks.
We implement all algorithms and evaluate their performances using Python language (3K+ lines of code).
All evaluations are conducted in a computer with two CPUs (Intel Xeon E5-2678 v3), one GPU (NVIDIA GeForce GTX 2080 Ti),
and 64GB memory, running Ubuntu 16.04.6 LTS.
All source code is publicly available in \texttt{https://github.com/DanzhouWu/TSRA}.

We first compare all our proposed random access algorithms, including the upper-bound algorithm, i.e., \eqref{equ:upper-bound-policy},
FSRA/HSRA/TSRA proposed in Sec.~\ref{sec:TSRA}, and the existing baseline, DLMA \cite{yiding2019deep}.
We simulate the deadline $D$ from 1 to 30. For each $D$,  we randomly select 500 groups of system parameters ($p_b$, $p_b'$, $p_s$, $p_s'$, $p_t$),
and independently run each group for 10,000,000 slots for FSRA and 100,000 slots  for the other four algorithms.
We then get the average performance of such 500 groups independently for the five algorithms.
The results are shown in Fig.~\ref{subfig:fig:timely_throu}. Note that the state spaces of the upper-bound algorithm and FSRA are of size
$2^{2D+2}$ and $2^{D+2}$, respectively, both of which increases exponentially with $D$.
Due to our computational resource limit, we can only evaluate the upper-bound algorithm for $D \le 5$,
and evaluate FSRA for $D \le 10$. Thus, we can see a truncation for both ``Upper Bound" and ``FSRA" curves in  Fig.~\ref{subfig:fig:timely_throu}.

From Fig.~\ref{subfig:fig:timely_throu}, we have the following three observations.
First, the upper bound proposed in Sec.~\ref{sec:upper_bound} indeed provides an effective means for evaluating the timely throughput of different algorithms. This holds by assuming the fact that user 2
has more revealed information, including user 1's parameters and queue information. In addition,
we can quantify the performance gap between the upper bound and any other algorithms.
For example, the system timely throughput of TSRA (resp. DLMA) is 4.98\% (resp. 10.83\%) less than that of the upper bound on
average for $D$ ranging from 1 to 5. Such a performance gap characterization was missing in many other works applying RL to network communication
problems \cite{yiding2019deep,yiding2020non,luong2019applications}. Second, TSRA has very close performance with HSRA and FSRA.
TSRA is only 0.50\% worse than FSRA on average for $D$ ranging from 1 to 10,
and only 0.15\% worse than HSRA on average for $D$ ranging from 1 to 30. This suggests that indeed
we can design the R-learning algorithm only depending on whether user 2 has a most urgent packet (whose lead time is 1).
Third, our proposed TSRA for delay-constrained heterogeneous wireless networks
achieves better performance than DLMA, which was designed for delay-unconstrained heterogeneous wireless networks.
The system timely throughput of TSRA is 5.62\% larger than that of DLMA on average for $D$ ranging from 1 to 30.

\begin{figure}[t]
  \centering
  \includegraphics[width=0.9\linewidth]{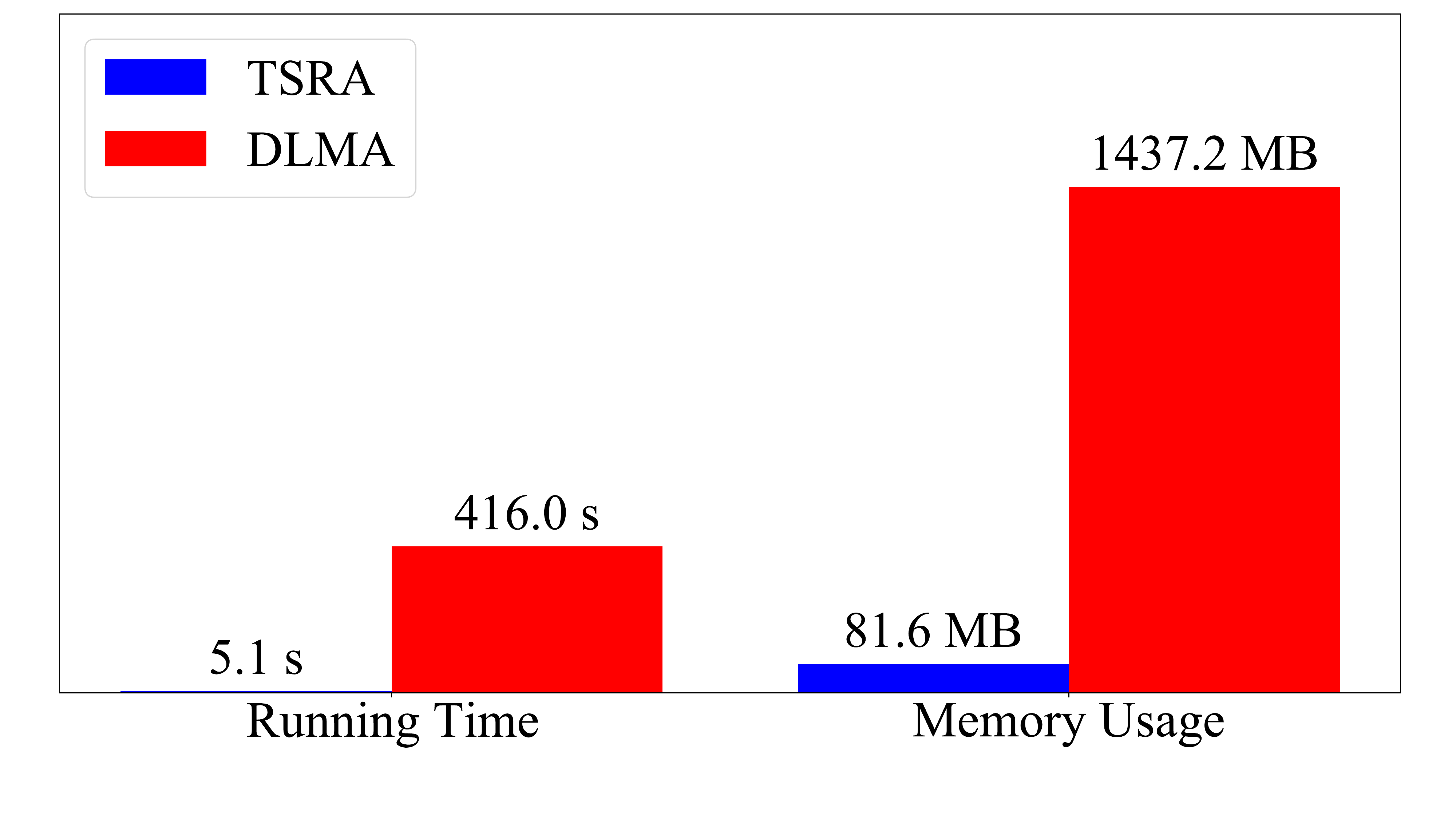}
  \caption{Comparison of the running times and memory usages of TSRA and DLMA where $p_b=0.5$, $p_b'=0.4$, $p_s=0.7$, $p_s'=0.6$, $p_t=0.4$, and $D=2$. \label{subfig:fig:time_memory} }
\end{figure}

In addition to the performance gain in terms of system timely throughput for TSRA over DLMA,
we further use Fig. \ref{subfig:fig:time_memory} to demonstrate that TSRA needs significantly less computational resource
than DLMA. We run one instance for TSRA and DLMA with $p_b=0.5$, $p_b'=0.4$, $p_s=0.7$, $p_s'=0.6$, $p_t=0.4$, and $D=2$.
The total number of running slots is $100,000$ for both algorithms. As we can see
from Fig. \ref{subfig:fig:time_memory}, TSRA only needs to run 5.1 seconds, over 80x less than that of DLMA,
and it only needs 81.6 MB of memory, over 17x less than that of DLMA. The reason is as follows.
In terms of time complexity, TSRA only needs to perform two simple computation steps (please refer to \eqref{equ:upgrade_Q} and \eqref{equ:upgrade_rho}) in each slot, while DLMA needs to go through a fully-connected multilayer neural network with significantly more computation operations in each slot.
In terms of space complexity, TSRA only needs to store the scalar $\rho$ and the Q-function table $Q(s,a)$,
where $s$ has only 8 possible values and $a$ has only 2 possible values (please refer to Sec.~\ref{sec:TSRA}).
However, DLMA needs to store a memory pool of 500 states, each of which is of size  $160$,
and the parameters of the fully-connected multilayer neural network \cite[Table 1]{yiding2019deep}.

\begin{figure}[t]
  \centering
  \includegraphics[width=0.9\linewidth]{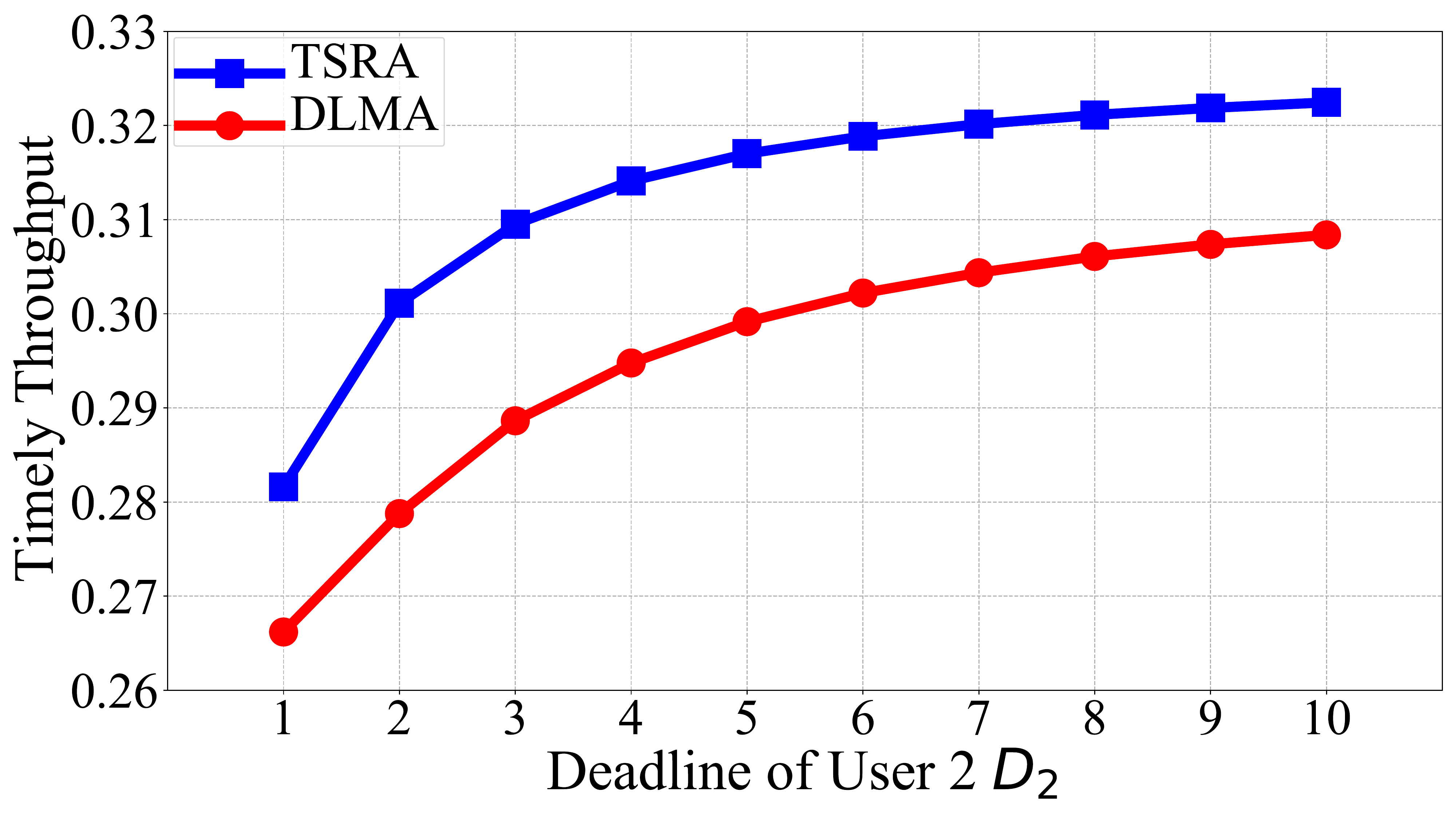}
  \caption{Comparison of the system timely throughputs of TSRA and DLMA when user 1 and user 2 have Bernoulli arrivals. We set $D_1=5$ and vary $D_2$ from 1 to 10. \label{fig:robustness-1} }
\end{figure}

\begin{figure}[t]
  \centering
  \includegraphics[width=0.9\linewidth]{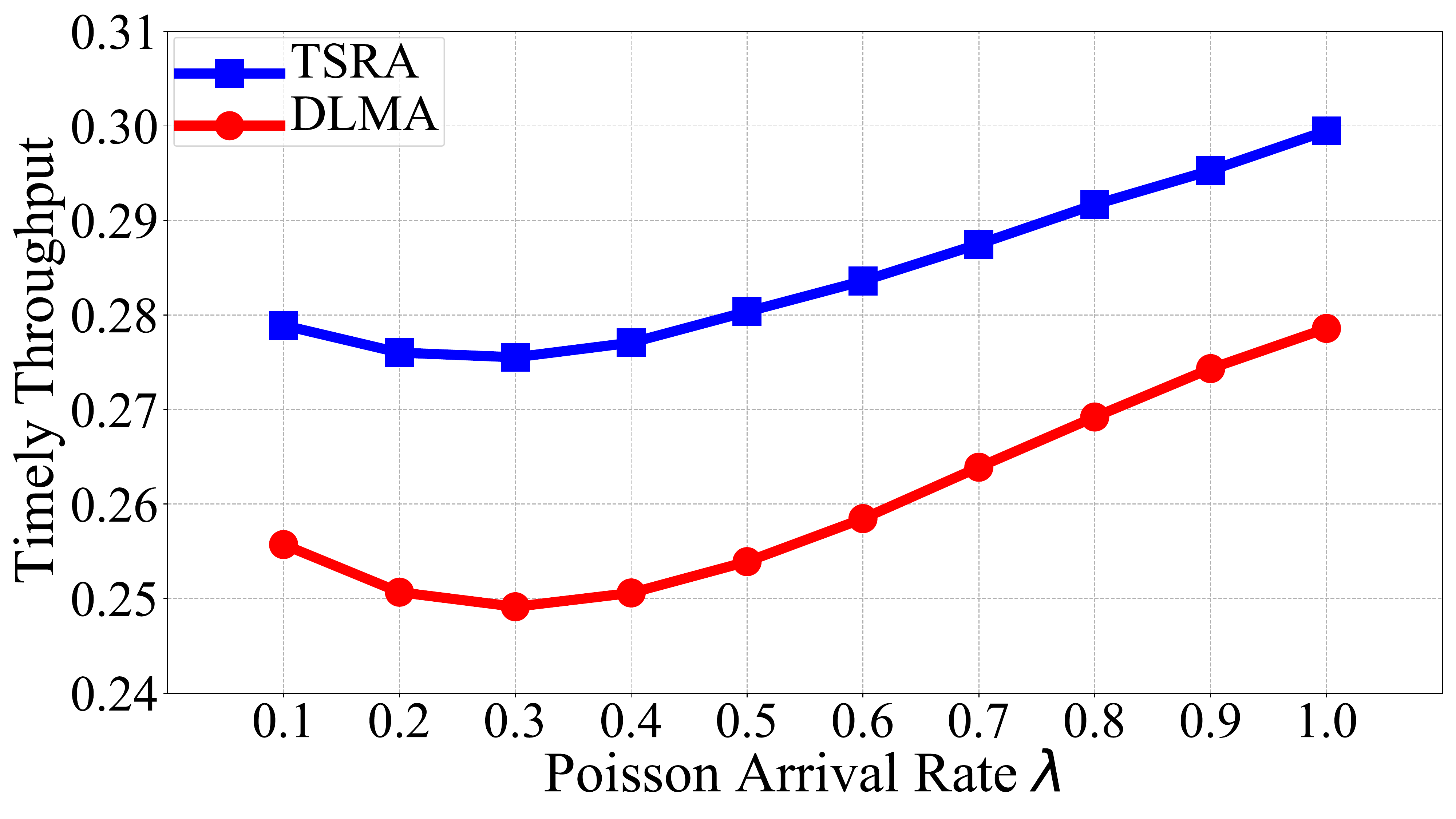}
  \caption{Comparison of the system timely throughputs of TSRA and DLMA when user 1 has Poisson arrivals with $D_1=2$ and user 2 has  Bernoulli arrivals with $D_2=2$. We vary Poisson arrival rate $\lambda$ from 0.1 to 1.0. \label{fig:robustness-2} }
\end{figure}

\begin{figure}[t]
  \centering
  \includegraphics[width=0.9\linewidth]{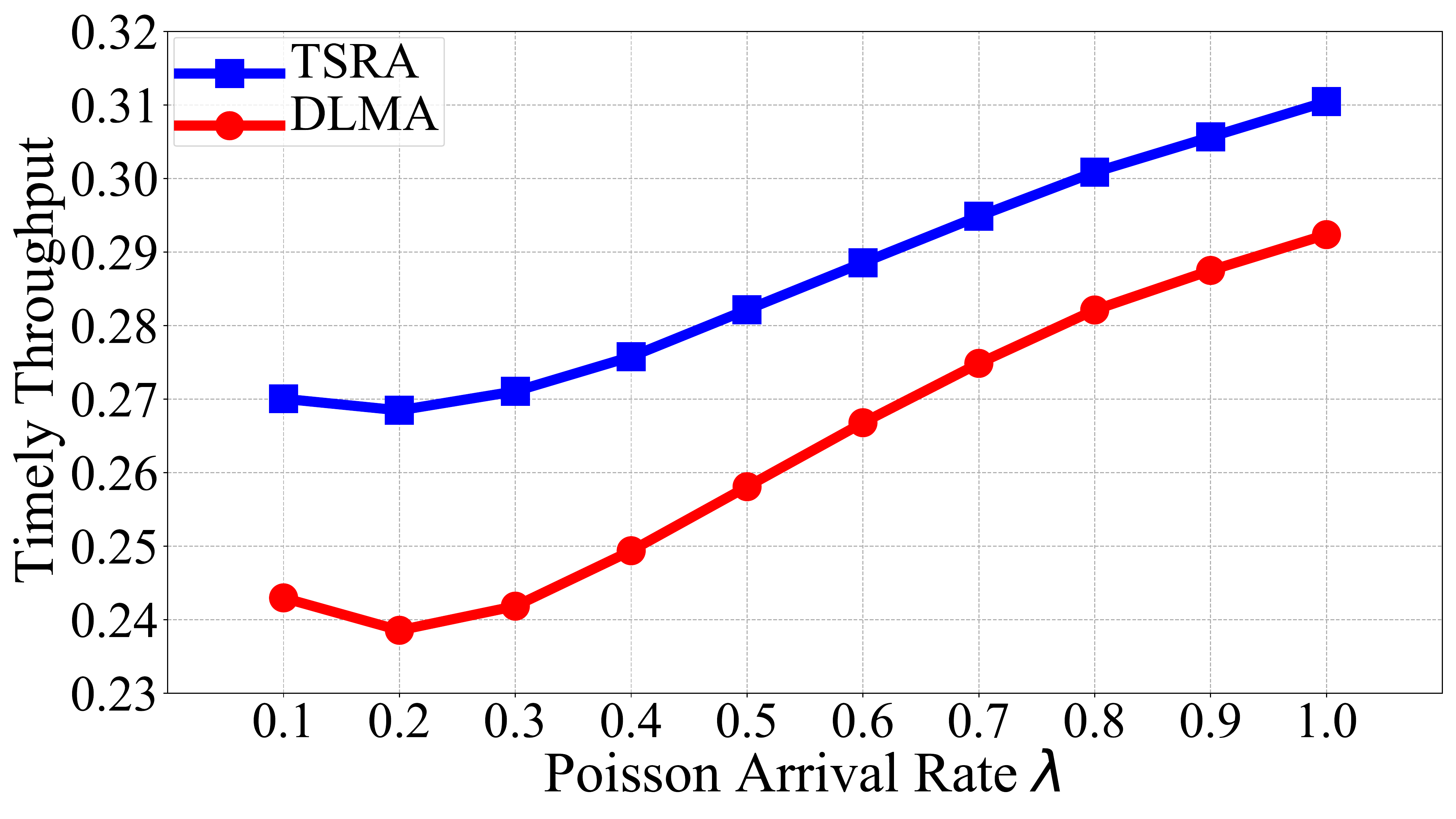}
  \caption{Comparison of the system timely throughputs of TSRA and DLMA when user 1 has Poisson arrivals with $D_1=4$ and user 2 has  Bernoulli arrivals with $D_2=2$. We vary Poisson arrival rate $\lambda$ from 0.1 to 1.0. \label{fig:robustness-3} }
\end{figure}

Finally, we demonstrate the robustness of our proposed TSRA algorithm for delay-constrained heterogeneous wireless networks.
In this paper, we assume that both users have Bernoulli arrivals and all their packets have the same deadline $D$.
We then consider three different settings with larger heterogeneity:
\begin{itemize}
\item Case 1 (Different deadlines): Both user 1 and user 2 have Bernoulli arrivals, but they have different deadline $D$'s  (Fig.~\ref{fig:robustness-1});
\item Case 2 (Different traffic patterns): User 1 has Poisson arrivals while user 2 has Bernoulli arrivals, but they have the same deadline (Fig.~\ref{fig:robustness-2});
\item Case 3 (Different traffic patterns and different deadlines): User 1 has Poisson arrivals with deadline $D_1$, while user 2 has Bernoulli arrivals with a different deadline $D_2$ (Fig.~\ref{fig:robustness-3}).
\end{itemize}
Note that for each point in Figs.~\ref{fig:robustness-1}-\ref{fig:robustness-3},
we get the average among randomly selected 500 groups of system parameters $(p_s,p_t,p'_b,p'_s)$, each run 100,000 slots.
We can observe that TSRA is again better than DLMA for all three cases.
On average, the system timely throughput of TSRA is 5.85\% more than that of DLMA in Fig.~\ref{fig:robustness-1},
9.30\% more that that of DLMA in Fig.~\ref{fig:robustness-2}, and 9.02\% more that that of DLMA in Fig.~\ref{fig:robustness-3}.
These results show that our proposed TSRA is robustly better than DLMA for different heterogeneous settings.

\begin{figure*}[t]
  \centering
  \subfigure[]{
    \label{fig:D-5-a-3} 
    \includegraphics[width=0.32\linewidth]{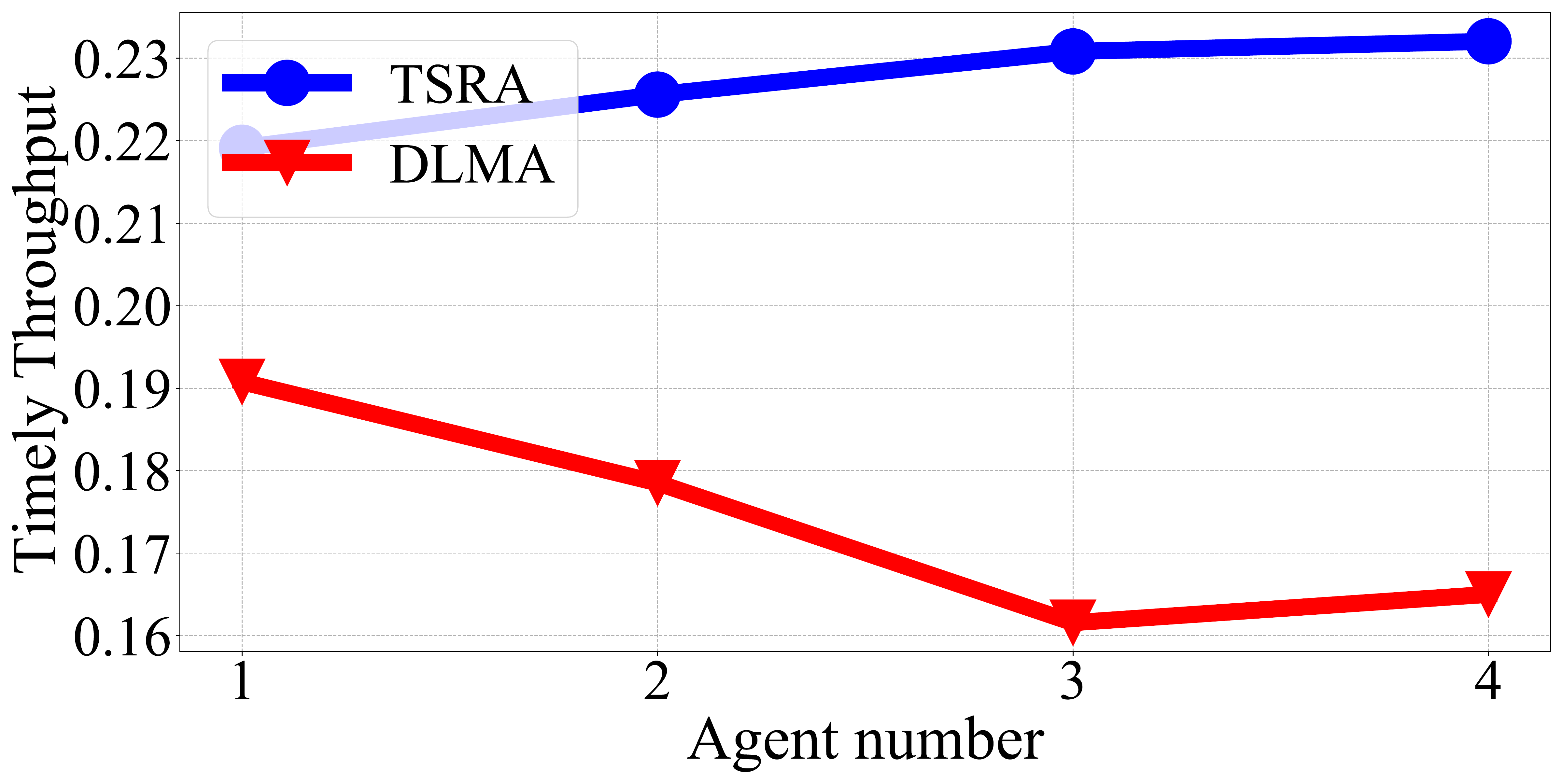}}
    \hfill
    \subfigure[]{
    \label{fig:D-7-a-2}
    \includegraphics[width=0.32\linewidth]{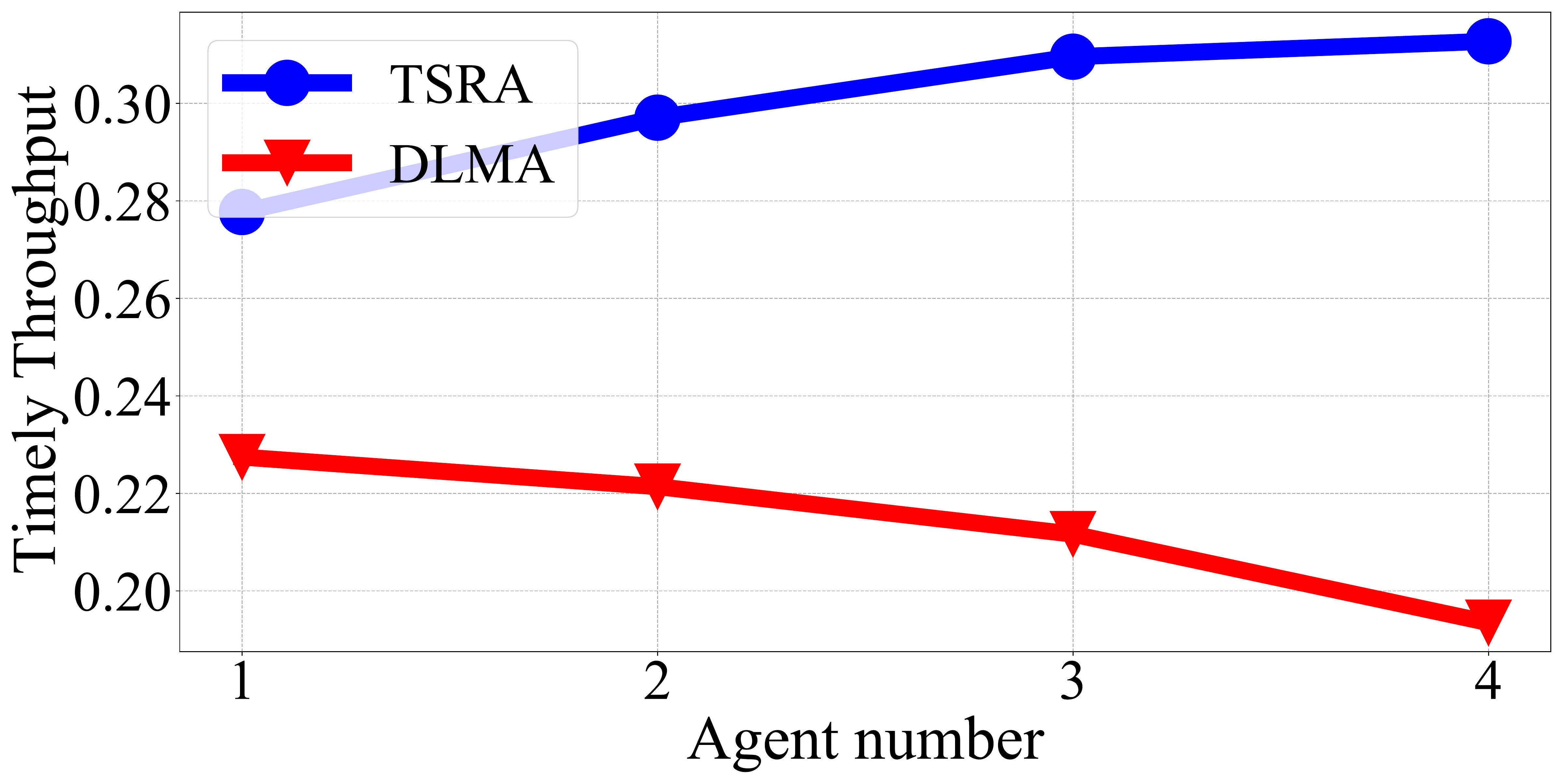}}
    \hfill
    \subfigure[]{
    \label{fig:D-10-a-1}
     \includegraphics[width=0.32\linewidth]{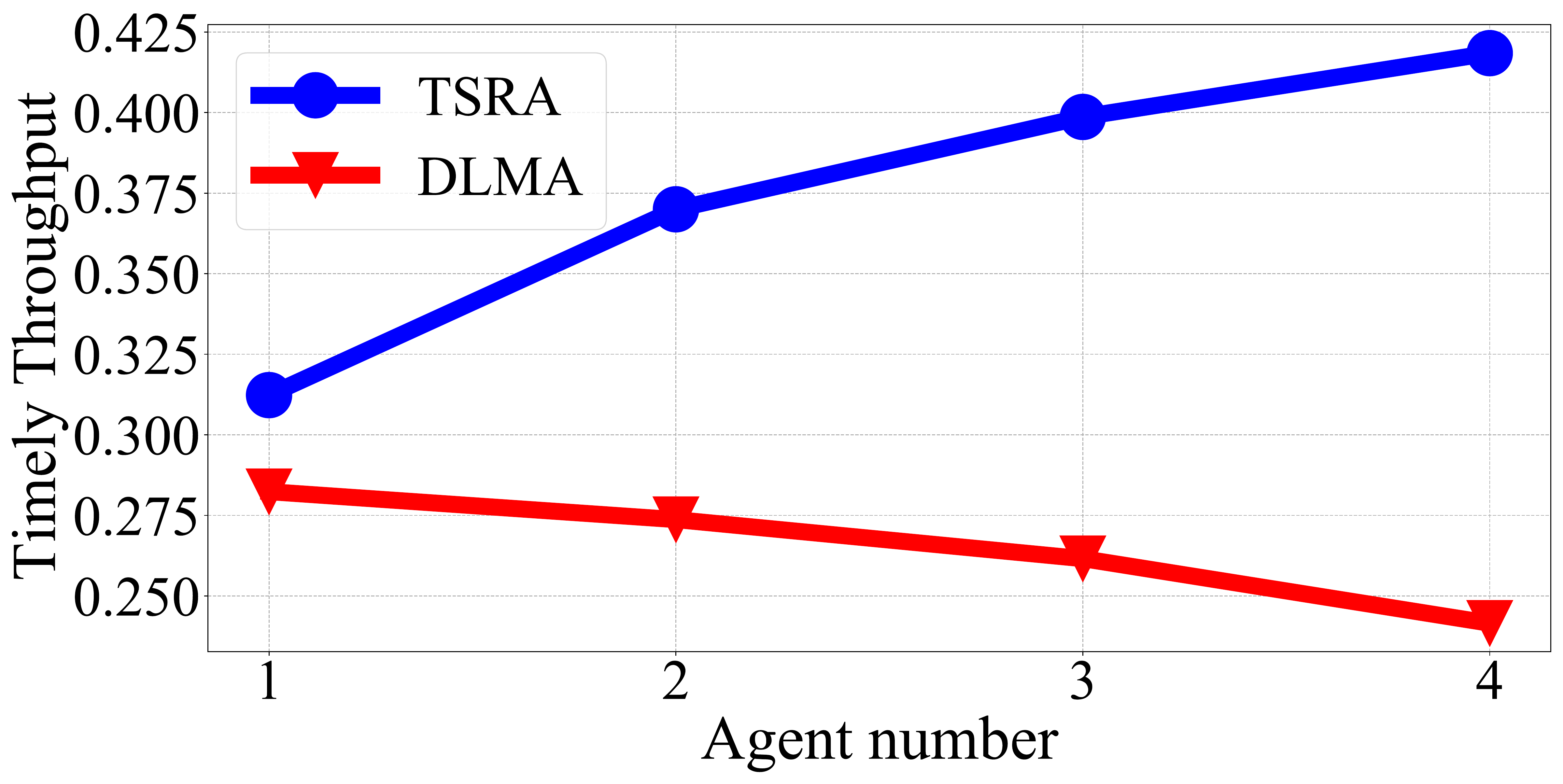}}
  \caption{Simulation results for multi-user case. (a)  $D=5$ and the number of ALOHA users is 3.
  (b) $D=7$ and the number of ALOHA users is 2.
  (c) $D=10$ and the number of ALOHA users is 1.
  \label{fig:multi-user-case-results} } 
\end{figure*}

In this paper,
as a first attempt to study the random access problem for a delay-constrained heterogeneous wireless network,
we consider a two-user case. We remark that a comprehensive study of multi-user case is beyond
the scope of this paper. However, to illustrate some first-order understandings,
we also simulate some multi-user cases and compare our proposed TSRA and the existing DLMA schemes.
We compared different deadlines $D$, different aloha numbers, and different reinforcement learning-based agents.
we randomly select 100 groups of system parameters, and independently run each group 100,000 slots for TSRA and DLMA.
We get the average performance of such 100 group independently for each setting.
The result is shown in Fig.~\ref{fig:multi-user-case-results}.
As we can see, our proposed TSRA also outperforms DLMA in such multi-user cases.

%% file: conclusion.tex
\section{Conclusion}\label{sec:conclusion}
In this paper, we for the first time investigate the random access problem for delay-constrained heterogeneous wireless networks.
We propose an R-learning-based low-complexity algorithm, called TSRA, for a two-user heterogeneous wireless network.
We show that TSRA achieves close-to-upper-bound performance and has better performance than the existing baseline DLMA \cite{yiding2019deep},
which was designed for delayed-unconstrained heterogeneous wireless networks.

Three key messages have been delivered by this work: 
First, although RL has been widely used in many network decision problems,
few works characterize their performance gap  due to  RL's black-box nature.
In this work, we instead propose an MDP-based formulation to derive a model-based upper bound
such that it can quantify the performance gap of any RL-based scheme. We believe that this methodology can
benefit other network problems utilizing RL. Second, since network problems are concerned with throughput or timely throughput, which is by nature an average reward, it is revealed by this work that
average-reward-based R-learning is better than the currently widely-used discounted-reward-based Q-learning.
Finally, for delay-constrained communications,
we show that the HoL packets or even the most urgent packets have the biggest impact on the system performance,
which can be utilized to simplify the system design significantly.

For future research of this ongoing work, it is interesting and important to study the random access problem for multi-user delay-constrained heterogeneous wireless networks.
In addition, it is also worthy to consider the user fairness, whilst maximizing the system timely throughput.

%% file: appendix.tex
\begin{appendix}

\subsection{Proof of Theorem~\ref{thm:D=1}} \label{app:proof-of-theorem-D=1}
For $D=1$, as we explained in Sec.~\ref{sec:upper_bound}, all slots are decoupled such
that we only need to focus on one particular slot. Thus,
we can shrink our design space to a single parameter, i.e., the transmission probability of user 2, which is denoted by $p'_t$.
To optimize $p'_t$, we first define the following random variables:
\begin{itemize}
\item Random variable $A_i, i=1,2$:
\be
A_i=
\left\{
     \begin{array}{ll}
       1, & \hbox{user $i$ has a non-expired packet to transmit;} \\
       0, & \hbox{otherwise.}
     \end{array}
   \right.
\ee
\item Random variable $X_i, i=1,2$:
\be
X_i=
\left\{
     \begin{array}{ll}
       1, & \hbox{user $i$ transmits a packet;} \\
       0, & \hbox{otherwise.}
     \end{array}
   \right.
\ee
\item Random variable $Y_i, i=1,2$:
\be
Y_i=
\left\{
     \begin{array}{ll}
       1, & \hbox{user $i$ transmits a packet successfully;} \\
       0, & \hbox{otherwise.}
     \end{array}
   \right.
\ee
\end{itemize}

Then, given parameters $p_b, p'_b, p_s, p'_s, p_t, p'_t$, we can derive the
distributions for the above random variables. The system timely throughput is,
\bee
R & = \mathbb{E} [Y_1 + Y_2] = \mathbb{E}[Y_1] + \mathbb{E}[Y_2] \nnb \\
& = P(Y_1=1) + P(Y_2=1). \label{equ:P-Y1=1+Y2=1}
\eee

Next we compute $P(Y_1=1)$ as follows,
\bee
& P(Y_1=1) \nnb \\
& \resizebox{.99\hsize}{!}{$ = \sum_{x_1 \in \{0,1\}} \sum_{x_2 \in \{0,1\}} P(Y_1=1|X_1=x_1, X_2=x_2) P(X_1=x_1, X_2=x_2)$} \nnb \\
&=P(Y_1=1|X_1=1, X_2=0)P(X_1=1, X_2=0)   \label{equ:Y1=1-e1} \\
&= p_s \cdot P(X_1=1, X_2=0) \label{equ:Y1=1-e2} \\
& = p_s \cdot P(X_1=1) \cdot P(X_2=0) \label{equ:Y1=1-e3},
\eee
where \eqref{equ:Y1=1-e1} holds because user 1 can deliver a packet successfully only if
user 1 transmits a packet and user 2 does not transmits a packet in the considered slot,
and \eqref{equ:Y1=1-e2} holds because the transmission events of both users are independent.

Now let us again use the law of total probability to  compute $P(X_1=1)$ and $P(X_2=0)$ in \eqref{equ:Y1=1-e3},
\bee
 P(X_1=1) & =   \sum_{a_1 \in \{0,1\}} P(X_1=1|A_1=a_1)P(A_1=a_1)  \nnb \\
& = P(X_1=1|A_1=1) P(A_1=1) \nnb \\
& = p_t p_b,  \label{equ:P-X1=1}
\eee
\bee
& P(X_2=0) =  \sum_{a_2 \in \{0,1\}} P(X_2=0|A_2=a_2)P(A_2=a_2) \nnb \\
& \resizebox{.99\hsize}{!}{$ = P(X_2=0|A_2=0) P(A_2=0) + P(X_2=0|A_2=1) P(A_2=1)$}\nnb \\
& = 1 \cdot (1-p'_b) + (1-p'_t) \cdot p'_b \nnb \\
& = 1 - p'_bp'_t. \label{equ:P-X2=0}
\eee

Inserting \eqref{equ:P-X1=1} and \eqref{equ:P-X2=0} into \eqref{equ:Y1=1-e3}, we obtain that
\be
P(Y_1=1) = p_sp_tp_b(1-p'_tp'_b). \label{equ:P-Y1=1}
\ee

Similarly, we can obtain
\be
P(Y_2=1) = p'_sp'_tp'_b(1-p_tp_b). \label{equ:P-Y2=1}
\ee

Inserting \eqref{equ:P-Y1=1} and \eqref{equ:P-Y2=1} into \eqref{equ:P-Y1=1+Y2=1}, we obtain the system
timely throughput as,
\bee
R &= p_sp_tp_b(1-p'_tp'_b) + p'_sp'_tp'_b(1-p_tp_b) \nnb \\
& = \left[ p'_{s}-(p_{s}+p'_{s})p_{t}p_{b} \right] p'_{t}p'_{b}+p_{s}p_{t}p_{b}. \nnb
\eee
Thus, if
\be
p_t p_b < \frac{p'_s}{p_s + p'_s}, \label{equ:binary-condition}
\ee
the optimal $p'_t$ to maximize the system timely throughput $R$ is
\be
p'_t=1,
\ee
i.e., user 2 will always transmit its packet if it has one packet.
Otherwise, if \eqref{equ:binary-condition} does not hold,
the optimal $p'_t$ to maximize the system timely throughput $R$ is
\be
p'_t=0,
\ee
i.e., user 2 will never transmit its packet.
This completes the proof.


\subsection{Why is FSQA worse than FSRA and how to improve FSQA?} \label{app:how_to_improve_FSQA}
\begin{figure}[t]
  \centering
  \includegraphics[width=0.9\linewidth]{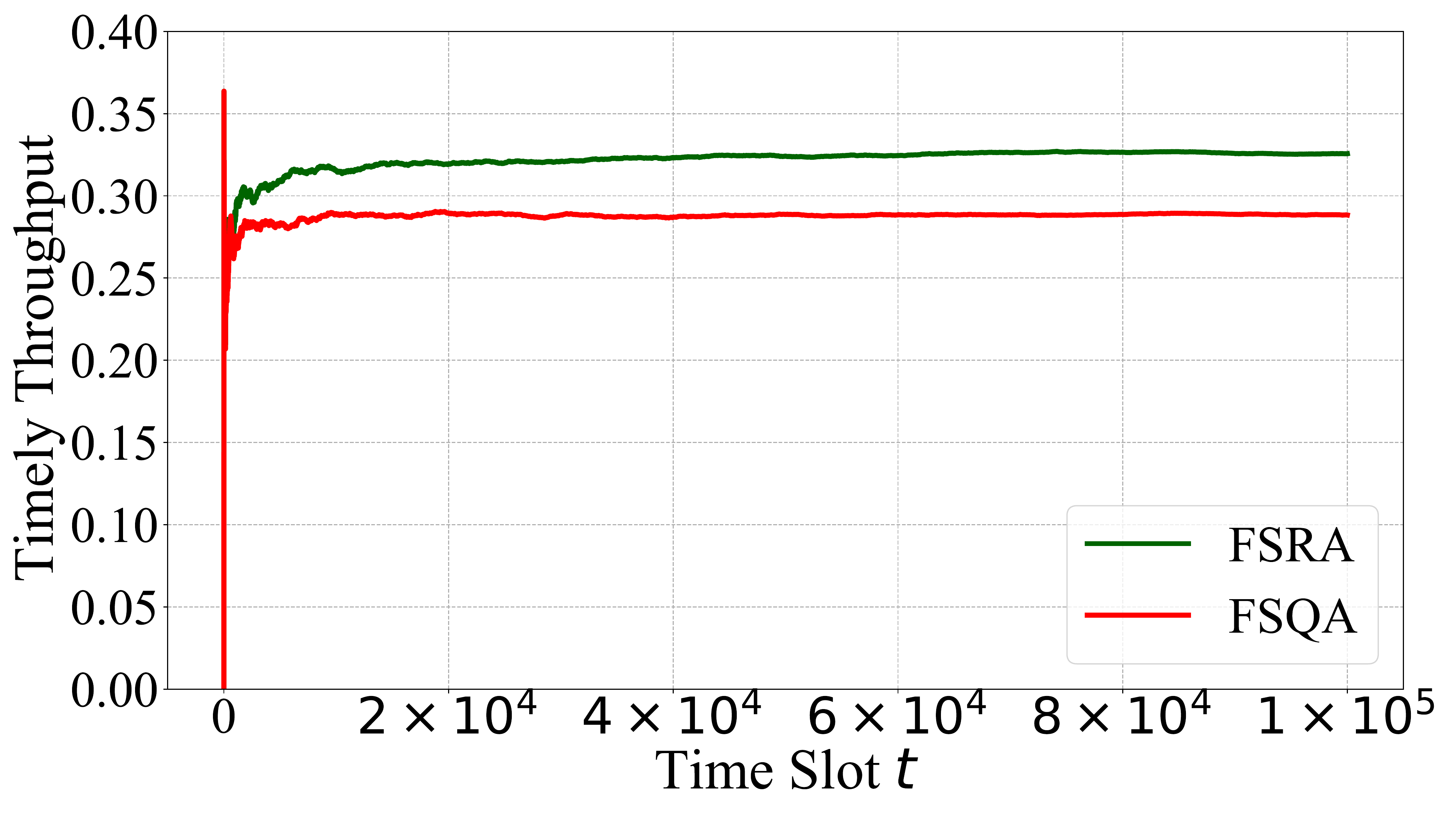}
  \caption{Compare the timely throughput of FSRA and FSQA where $p_b=0.5$, $p_b'=0.4$, $p_s=0.7$, $p_s'=0.6$, $p_t=0.4$, $D=2$.}\label{fig:performance_ql_rl}
\end{figure}

As we showed in Fig.~\ref{fig:compare_rl_ql}, the Q-learning-based FSQA algorithm
is worse than the R-learning-based FSRA. In this part, we consider a specific example
to delve into the details of FSQA and FSRA.
We set system parameter settings as $p_b=0.5$, $p_b'=0.4$, $p_s=0.7$, $p_s'=0.6$, $p_t=0.4$, $D=2$.
The achieved system timely throughput of FSQA and FSRA is show in Fig.~\ref{fig:performance_ql_rl}.
Obviously, FSRA outperforms FSQA. Now we take a further step to examine the random access policies
of FSQA and FSRA, which are shown in Table~\ref{tab:FSRA-FSQA-policy}. As we can see, indeed, after convergence,
FSQA ahd FSRA take different policies, which thus results in different system timely throughput.

It is not clear which policy is better. We then use the upper-bound policy as a benchmark, i.e., \eqref{equ:upper-bound-policy},
to justify that the policy of FSRA is better. Note that in the upper-bound algorithm, we
use a model-based MDP formulation where user 2 is aware of user 1's queue information and parameters.
Thus, different from FSQA and FSRA whose system state is $s=(l_2,o)$ as shown in \eqref{equ:equ-state-FSQA-and-FSRA},
the system state of the upper-bound algorithm also includes user 1's queue information, i.e., $s=(l_1,l_2,o)$,
as shown in \eqref{equ:S1}.  The policy of the upper-bound algorithm is shown in Table~\ref{tab:upper-bound-policy}.
Each state of FSQA and FSRA, i.e., $s=(l_2,o)$, corresponds to four states of upper-bound policy, i.e, $s=(l_1,l_2,o)$
where $l_1 \in \{(0,0),(0,1),(1,0),(1,1)\}$. For such four states of the upper-bound policy sharing the same $l_2$ and $o$,
we take a vote to obtain the majority action, which is the last column in Table~\ref{tab:upper-bound-policy}. The majority action
roughly represents the optimal action if the user 2's queue information is $l_2$ and the channel observation is $o$.
We compare the majority action of the upper-bound policy in Table \ref{tab:upper-bound-policy} and the action of FSRA and FSQA in Table~\ref{tab:FSRA-FSQA-policy}.
We can see that FSRA has exactly the same action with the upper-bound policy for all states, while FSQA has different actions for
four states $(l_2,o)=((1,0), \textsf{SUCCESSFUL})$, $((1,1), \textsf{BUSY})$, $((1,1), \textsf{SUCCESSFUL}),$ and $((1,1), \textsf{FAILED})$.
With the help of the model-based upper-bound policy as a benchmark, we can see that indeed the policy of R-learning-based FSRA is better than
the policy of Q-learning-based FSQA.

Furthermore, we also use this example to show how to improve the performance of Q-learning-based FSQA algorithm.
Comparing the Q-function update of Q-learning in \eqref{equ:Q-learning-Q-function} and
the Q-function update of R-learning in \eqref{equ:upgrade_Q}, we can see that
the major difference is the parameter $\rho$.
Comparing \eqref{equ:Q-function-approx} and \eqref{equ:Q-approx-R-learning},
which respectively represents the physical meaning of Q-function for Q-learning and R-learning,
we can also observe that for average-reward MDP, we should use a relative value to response the reward. Namely,
the reward should be deducted by a constant $\rho$. To improve the Q-learning-based FSQA algorithm,
we thus re-define its reward function in \eqref{equ:reward function_RL} as
\bee
r(s_t,a_t) \triangleq 1_{\left\{o_t \in \{ \textsf{BUSY}, \textsf{SUCCESSFUL}\} \right \}}-c, \forall s_t \in \mathcal{S}', a_t \in \mathcal{A},
\label{equ:reward function_RL-redefined}
\eee
where $c \in [0,1]$ is a constant. We then compare the performance of FSRA and the improved FSQA algorithms with different $c$'s, as shown
in Fig.~\ref{fig:performance_ql_rl}. As we can see, when constant $c=0.3$, the improved FSQA achieves almost the same system timely throughput
with FSRA, which is much better the original FSQA algorithm (with $c=0$). In fact, parameter $\rho$ in \eqref{equ:upgrade_rho} of FSRA converges to 0.379 in this example.
Thus, the optimal constant $c=0.3$ in the improved FSQA is close to the converged $\rho$ of FSRA.
Although we can improve FSQA by re-defining its reward function according to \eqref{equ:reward function_RL-redefined},
there is generally no guidance on how to choose the best constant $c$, which is different for different problem instances.
Instead, in R-learning-based FSRA, the parameter $\rho$ is algorithmically  adjusted according to \eqref{equ:upgrade_rho} until its convergence.
This further demonstrates the benefit of R-learning over Q-learning for our studied problem.

\begin{figure}[t]
  \centering
  \includegraphics[width=0.9\linewidth]{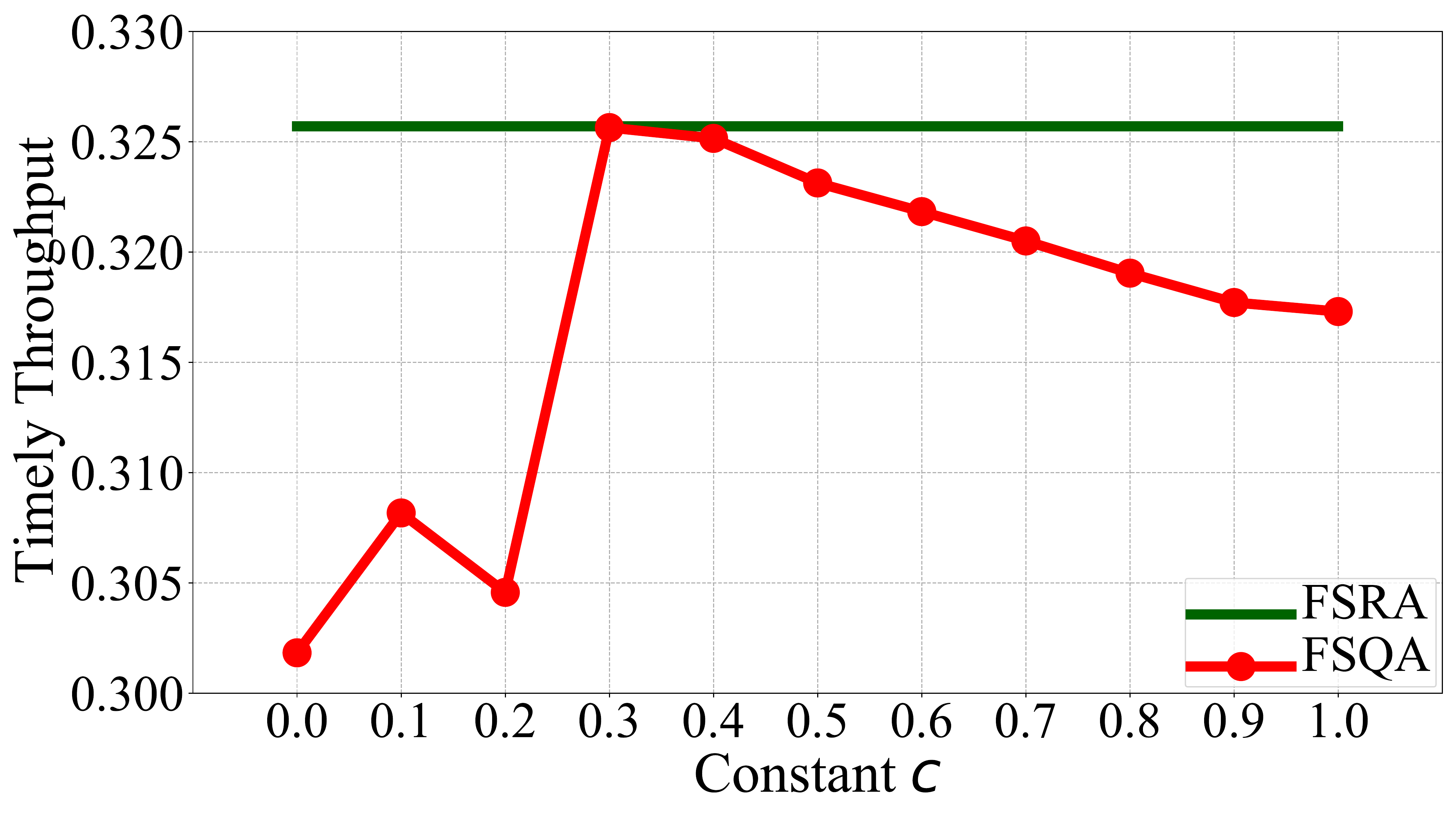}
  \caption{Compare the timely throughput of FSRA and the improved FSQA where $p_b=0.5$, $p_b'=0.4$, $p_s=0.7$, $p_s'=0.6$, $p_t=0.4$, $D=2$.}\label{fig:constant}
\end{figure}

\begin{table}[]
\begin{center}
\caption{The policies of FSRA and FSQA after convergence when $p_b=0.5$, $p_b'=0.4$, $p_s=0.7$, $p_s'=0.6$, $p_t=0.4$ and $D=2$.
Note that channel observation $o=\text{B}$ means that $o=\textsf{BUSY}$,
$o=\text{S}$ means that $o=\textsf{SUCCESSFUL}$, $o=\text{I}$ means that $o=\textsf{IDLE}$,
and $o=\text{F}$ means that $o=\textsf{FAILED}$. \label{tab:FSRA-FSQA-policy}}
\begin{tabular}{|c|c|c|c|}
\hline
\multicolumn{2}{|c|}{State}     & FSRA         & FSQA         \\ \hline
\multicolumn{2}{|c|}{$s=(l_2, o)$} & \multicolumn{2}{c|}{action} \\ \hline
$l_2$               & $o$            & \multicolumn{2}{c|}{$a$}      \\ \hline
(0,0)            & B            & \textsf{WAIT}         & \textsf{WAIT}         \\ \hline
(0,0)            & S            & \textsf{WAIT}         & \textsf{WAIT}         \\ \hline
(0,0)            & I            & \textsf{WAIT}         & \textsf{WAIT}         \\ \hline
(0,0)            & F            & \textsf{WAIT}         & \textsf{WAIT}         \\ \hline
(0,1)            & B            & \textsf{TRANSMIT}     & \textsf{TRANSMIT}     \\ \hline
(0,1)            & S            & \textsf{TRANSMIT}     & \textsf{TRANSMIT}     \\ \hline
(0,1)            & I            & \textsf{TRANSMIT}     & \textsf{TRANSMIT}     \\ \hline
(0,1)            & F            & \textsf{TRANSMIT}     & \textsf{TRANSMIT}     \\ \hline
(1,0)            & B            & \textsf{TRANSMIT}     & \textsf{TRANSMIT}     \\ \hline
(1,0)            & S            & \textsf{TRANSMIT}     & \textsf{WAIT}         \\ \hline
(1,0)            & I            & \textsf{TRANSMIT}     & \textsf{TRANSMIT}     \\ \hline
(1,0)            & F            & \textsf{TRANSMIT}     & \textsf{TRANSMIT}     \\ \hline
(1,1)            & B            & \textsf{TRANSMIT}     & \textsf{WAIT}         \\ \hline
(1,1)            & S            & \textsf{TRANSMIT}     & \textsf{WAIT}         \\ \hline
(1,1)            & I            & \textsf{TRANSMIT}     & \textsf{TRANSMIT}     \\ \hline
(1,1)            & F            & \textsf{TRANSMIT}     & \textsf{WAIT}         \\ \hline
\end{tabular}
\end{center}
\end{table}

\begin{table}[]
\begin{center}
\caption{The upper-bound policy \eqref{equ:upper-bound-policy} when $p_b=0.5$, $p_b'=0.4$, $p_s=0.7$, $p_s'=0.6$, $p_t=0.4$ and $D=2$.
Note that channel observation $o=\text{B}$ means that $o=\textsf{BUSY}$,
$o=\text{S}$ means that $o=\textsf{SUCCESSFUL}$, $o=\text{I}$ means that $o=\textsf{IDLE}$,
and $o=\text{F}$ means that $o=\textsf{FAILED}$.\label{tab:upper-bound-policy}}
\begin{tabular}{|c|c|c|c|c|c|}
\hline
\multicolumn{3}{|c|}{State}                         & \multicolumn{2}{c|}{$\pi(s|a)$} & \multirow{3}{*}{Majority} \\ \cline{1-5}
\multicolumn{3}{|c|}{$s=(l_1, l_2, o)$}                 & \multicolumn{2}{c|}{$a$}       &                           \\ \cline{1-5}
$l_1$    & $l_2$                     & $o$                  & \textsf{WAIT}         & \textsf{TRANSMIT}      &                           \\ \hline
(0,0) & \multirow{4}{*}{(0,0)} & \multirow{4}{*}{B} & 1            & 0             & \multirow{4}{*}{\textsf{WAIT}}     \\ \cline{1-1} \cline{4-5}
(0,1) &                        &                    & 1            & 0             &                           \\ \cline{1-1} \cline{4-5}
(1,0) &                        &                    & 1            & 0             &                           \\ \cline{1-1} \cline{4-5}
(1,1) &                        &                    & 1            & 0             &                           \\ \hline
(0,0) & \multirow{4}{*}{(0,0)} & \multirow{4}{*}{S} & 1            & 0             & \multirow{4}{*}{\textsf{WAIT}}     \\ \cline{1-1} \cline{4-5}
(0,1) &                        &                    & 1            & 0             &                           \\ \cline{1-1} \cline{4-5}
(1,0) &                        &                    & 1            & 0             &                           \\ \cline{1-1} \cline{4-5}
(1,1) &                        &                    & 1            & 0             &                           \\ \hline
(0,0) & \multirow{4}{*}{(0,0)} & \multirow{4}{*}{I} & 1            & 0             & \multirow{4}{*}{\textsf{WAIT}}     \\ \cline{1-1} \cline{4-5}
(0,1) &                        &                    & 1            & 0             &                           \\ \cline{1-1} \cline{4-5}
(1,0) &                        &                    & 1            & 0             &                           \\ \cline{1-1} \cline{4-5}
(1,1) &                        &                    & 1            & 0             &                           \\ \hline
(0,0) & \multirow{4}{*}{(0,0)} & \multirow{4}{*}{F} & 1            & 0             & \multirow{4}{*}{\textsf{WAIT}}     \\ \cline{1-1} \cline{4-5}
(0,1) &                        &                    & 1            & 0             &                           \\ \cline{1-1} \cline{4-5}
(1,0) &                        &                    & 1            & 0             &                           \\ \cline{1-1} \cline{4-5}
(1,1) &                        &                    & 1            & 0             &                           \\ \hline
(0,0) & \multirow{4}{*}{(0,1)} & \multirow{4}{*}{B} & 0            & 1             & \multirow{4}{*}{\textsf{TRANSMIT}} \\ \cline{1-1} \cline{4-5}
(0,1) &                        &                    & 0            & 1             &                           \\ \cline{1-1} \cline{4-5}
(1,0) &                        &                    & 1            & 0             &                           \\ \cline{1-1} \cline{4-5}
(1,1) &                        &                    & 0            & 1             &                           \\ \hline
(0,0) & \multirow{4}{*}{(0,1)} & \multirow{4}{*}{S} & 0            & 1             & \multirow{4}{*}{\textsf{TRANSMIT}} \\ \cline{1-1} \cline{4-5}
(0,1) &                        &                    & 0            & 1             &                           \\ \cline{1-1} \cline{4-5}
(1,0) &                        &                    & 1            & 0             &                           \\ \cline{1-1} \cline{4-5}
(1,1) &                        &                    & 0            & 1             &                           \\ \hline
(0,0) & \multirow{4}{*}{(0,1)} & \multirow{4}{*}{I} & 0            & 1             & \multirow{4}{*}{\textsf{TRANSMIT}} \\ \cline{1-1} \cline{4-5}
(0,1) &                        &                    & 0            & 1             &                           \\ \cline{1-1} \cline{4-5}
(1,0) &                        &                    & 1            & 0             &                           \\ \cline{1-1} \cline{4-5}
(1,1) &                        &                    & 0            & 1             &                           \\ \hline
(0,0) & \multirow{4}{*}{(0,1)} & \multirow{4}{*}{F} & 0            & 1             & \multirow{4}{*}{\textsf{TRANSMIT}} \\ \cline{1-1} \cline{4-5}
(0,1) &                        &                    & 0            & 1             &                           \\ \cline{1-1} \cline{4-5}
(1,0) &                        &                    & 1            & 0             &                           \\ \cline{1-1} \cline{4-5}
(1,1) &                        &                    & 0            & 1             &                           \\ \hline
(0,0) & \multirow{4}{*}{(1,0)} & \multirow{4}{*}{B} & 0            & 1             & \multirow{4}{*}{\textsf{TRANSMIT}} \\ \cline{1-1} \cline{4-5}
(0,1) &                        &                    & 0            & 1             &                           \\ \cline{1-1} \cline{4-5}
(1,0) &                        &                    & 0.47916      & 0.52084       &                           \\ \cline{1-1} \cline{4-5}
(1,1) &                        &                    & 0.49551      & 0.50449       &                           \\ \hline
(0,0) & \multirow{4}{*}{(1,0)} & \multirow{4}{*}{S} & 0            & 1             & \multirow{4}{*}{\textsf{TRANSMIT}} \\ \cline{1-1} \cline{4-5}
(0,1) &                        &                    & 0            & 1             &                           \\ \cline{1-1} \cline{4-5}
(1,0) &                        &                    & 0            & 1             &                           \\ \cline{1-1} \cline{4-5}
(1,1) &                        &                    & 0            & 1             &                           \\ \hline
(0,0) & \multirow{4}{*}{(1,0)} & \multirow{4}{*}{I} & 0            & 1             & \multirow{4}{*}{\textsf{TRANSMIT}} \\ \cline{1-1} \cline{4-5}
(0,1) &                        &                    & 0            & 1             &                           \\ \cline{1-1} \cline{4-5}
(1,0) &                        &                    & 0.35845      & 0.64155       &                           \\ \cline{1-1} \cline{4-5}
(1,1) &                        &                    & 0.43222      & 0.56778       &                           \\ \hline
(0,0) & \multirow{4}{*}{(1,0)} & \multirow{4}{*}{F} & 0            & 1             & \multirow{4}{*}{\textsf{TRANSMIT}} \\ \cline{1-1} \cline{4-5}
(0,1) &                        &                    & 0            & 1             &                           \\ \cline{1-1} \cline{4-5}
(1,0) &                        &                    & 0            & 1             &                           \\ \cline{1-1} \cline{4-5}
(1,1) &                        &                    & 0            & 1             &                           \\ \hline
(0,0) & \multirow{4}{*}{(1,1)} & \multirow{4}{*}{B} & 0            & 1             & \multirow{4}{*}{\textsf{TRANSMIT}} \\ \cline{1-1} \cline{4-5}
(0,1) &                        &                    & 0            & 1             &                           \\ \cline{1-1} \cline{4-5}
(1,0) &                        &                    & 0.49072      & 0.50928       &                           \\ \cline{1-1} \cline{4-5}
(1,1) &                        &                    & 0.49713      & 0.50287       &                           \\ \hline
(0,0) & \multirow{4}{*}{(1,1)} & \multirow{4}{*}{S} & 0            & 1             & \multirow{4}{*}{\textsf{TRANSMIT}} \\ \cline{1-1} \cline{4-5}
(0,1) &                        &                    & 0            & 1             &                           \\ \cline{1-1} \cline{4-5}
(1,0) &                        &                    & 0            & 1             &                           \\ \cline{1-1} \cline{4-5}
(1,1) &                        &                    & 0            & 1             &                           \\ \hline
(0,0) & \multirow{4}{*}{(1,1)} & \multirow{4}{*}{I} & 0            & 1             & \multirow{4}{*}{\textsf{TRANSMIT}} \\ \cline{1-1} \cline{4-5}
(0,1) &                        &                    & 0            & 1             &                           \\ \cline{1-1} \cline{4-5}
(1,0) &                        &                    & 0.41202      & 0.58798       &                           \\ \cline{1-1} \cline{4-5}
(1,1) &                        &                    & 0.44799      & 0.55201       &                           \\ \hline
(0,0) & \multirow{4}{*}{(1,1)} & \multirow{4}{*}{F} & 0            & 1             & \multirow{4}{*}{\textsf{TRANSMIT}} \\ \cline{1-1} \cline{4-5}
(0,1) &                        &                    & 0            & 1             &                           \\ \cline{1-1} \cline{4-5}
(1,0) &                        &                    & 0            & 1             &                           \\ \cline{1-1} \cline{4-5}
(1,1) &                        &                    & 0            & 1             &                           \\ \hline
\end{tabular}
\end{center}
\end{table}

\end{appendix}